\documentclass{aa}  
\usepackage{natbib}
\bibpunct{(}{)}{;}{a}{}{,} 
\usepackage{graphicx}
\usepackage{txfonts}
\usepackage{blindtext}
\usepackage[colorlinks=true]{hyperref}
\usepackage{placeins}
\usepackage{tikz}
\usepackage[export]{adjustbox}
\usetikzlibrary{shapes,arrows}
\usepackage{mathastext}

\tikzstyle{block} = [rectangle,thick, draw=black!100, 
    text width=1.2em, text centered, rounded corners, minimum height=1em,minimum width = 0.6em]
\tikzstyle{block2} = [rectangle,thick, draw=black!100,  
    text width=2.4em, text centered, rounded corners, minimum height=1em,minimum width = 1.2em]
\tikzstyle{line} = [draw,ultra thick, -latex']
\tikzstyle{curved line} = [draw, bend right=45,ultra thick, -latex']
\tikzstyle{dashed line} = [draw,ultra thick, -latex',dashed]
\tikzstyle{cloud} = [thick, ellipse, draw=black!100,  
    node distance=0.6cm, minimum width=2em,
    minimum height=1.2em] 
    
\begin{document} 

\title{DEFROST: Detecting Excess in Faraday Rotation thrOugh Sophisticated analysis Techniques}
   \subtitle{}

   \author{
   Valentina Vacca
   \inst{1,2}
\and
Sebastian Hutschenreuter
\inst{3}
 \and
Andrea Cabriolu
\inst{1,2,4}
 \and
Torsten A. En{\ss}lin
\inst{2,5}
 \and
 Jakob Roth
 \inst{2,6}
\and
Martin Reineke
 \inst{2}
 \and
Philipp Frank
 \inst{2}
 \and
Federica Govoni
\inst{1}
\and
Matteo Murgia 
\inst{1}
\and 
 Gianni Fenu
 \inst{4}
             }

   \institute{
INAF-Osservatorio
Astronomico di Cagliari, Via della Scienza 5, I-09047 Selargius (CA), Italy
\and
Max Planck Institute for Astrophysics, Karl-Schwarzschildstr. 1, 85741 Garching, Germany
\and
University of Vienna, Department of Astrophysics, T{\"u}rkenschanzstrasse 17, 1180 Vienna, Austria
\and
Department of Mathematics and Computer Science, University of Cagliari, Via Ospedale 72, Cagliari, 09121, Italy
\and
Deutsches Zentrum für Astrophysik, Postplatz 1, 02826 Görlitz, Germany
\and
Technische Universität M\"unchen (TUM), Boltzmannstr. 3, 85748 Garching,
Germany}
   \date{Accepted for publication on 11 May 2026}

  \abstract
  {}
  {Understanding the origin and evolution of cosmological magnetic fields requires a detailed knowledge of the strength of magnetic fields in different extragalactic environments. In this context, a powerful tool is the statistical analysis of the Faraday effect on the linear polarization of a sample of radio sources. This effect carries information about the magnetic fields in our Galaxy, in extragalactic environments between the sources and the observer, and within the emitting radio source itself. An accurate disentangling of all these components is crucial to characterize magnetic fields in the large-scale structure of the Universe.}{The significant amount of data delivered by new radio instruments enables the investigation of increasingly weak magnetic fields. However, a trustworthy characterization is only possible with advanced analysis techniques. In this work, we present a new algorithm capable of simultaneously disentangling the Faraday effect due to our Galaxy from extragalactic contributions, by properly taking into account the observing noise. The algorithm takes as an input a catalog of Faraday depth measurements complemented by auxiliary information as, e.g., the redshift of the sources.}
    {We tested the algorithm with synthetic data to assess its performance and identify the range of slopes of the Galactic magnetic field power spectrum that allows us to properly disentangle Galactic and extragalactic contributions. Furthermore, we tested the algorithm with synthetic catalogs, based on meter and centimeter data currently available, corresponding to different observing setups, noise, and cuts in the absolute value of the Galactic latitude of the radio sources. Considering noise values and density of polarized sources consistent with existing catalogs, we demonstrated that the most robust results are obtained with sources with absolute Galactic latitude greater than 45 degrees, with inference of the extragalactic parameters at most within 5-$\sigma$, both for dispersion in Faraday rotation of about one and ten rad/m$^2$.}
    {}


   \maketitle

\section{Introduction}
\label{intro}

Magnetic fields control the acceleration and propagation of cosmic rays, contribute to the energetic balance of the system, and can deeply affect energy transfer and gas dynamics. Uncovering the origin and evolution of cosmological magnetic fields and quantifying their contribution from an energetic and dynamical point of view is one of the most challenging modern goals of extragalactic radio astronomy \citep[e.g., ][]{Heald2020}. 

Polarimetric radio observations enable the study of extragalactic magnetic fields via the Faraday effect, which causes a rotation of the polarization plane from background radio sources.
In the simpler configuration of an extragalactic emitting radio source in the background of a rotating magneto-ionic plasma, the Faraday rotation can be constrained by measuring the polarization angle $\Psi$ at two or more wavelengths $\lambda$,
\begin{equation}
\Psi=\Psi_0+RM\,\lambda^2,
\end{equation}
where $\Psi_0$ is the intrinsic polarization angle of the source and 
RM is the rotation measure defined by
\begin{equation}
RM \equiv \frac{\partial \Psi}{\partial \lambda^2}.
\end{equation}

When the radio emitting plasma is mixed with a rotating plasma, the observed polarization angle does not vary linearly as a function of $\lambda^2$ and more complex approaches are needed, such as rotation measure synthesis \citep{Brentjens2005}. To describe these scenarios, \cite{Burn1966} introduced the  Faraday depth,
\begin{equation}
\phi=0.81\int_{\textit{z}}^0\frac{d\textit{l}}{d\textit{z}}\textit{n(z)B}_{\rm l}(z)d\textit{z} ~~~~{\rm rad/m^2},
\end{equation}
where $\textit{n}$ is the thermal gas density in cm$^{-3}$, $\textit{B}_{\rm l}$ is the component of the magnetic field along the line of sight in $\mu$G, $\textit{z}$ is the redshift of the source, and $d\textit{l}$ is the infinitesimal path length in pc.

Statistical analysis of the Faraday effect on collections of polarized point sources, the so-called RM-grids, is a method for detecting and characterizing weak magnetic fields in low-density environments, the detection of which is extremely complex. Recently, Faraday depth values derived from several polarimetric surveys were used to produce a consolidated catalog of approximately sixty thousand measurements by \cite{VanEck2023}, mainly from the NRAO VLA Sky Survey \citep[NVSS, ][]{Condon1998,Taylor2009}. 
New upcoming catalogs, obtained with data from the Australian SKA Pathfinder (ASKAP), will dramatically increase this number: the Spectra and Polarisation in
Cutouts of Extragalactic Sources from the Rapid ASKAP Continuum Survey \citep[SPICE-RACS][Thomson in prep.]{Thomson2023} will provide an initial RM grid with a density of 7 polarized sources per square degree over 36,100\,deg$^2$, and the Polarisation Sky Survey of the Universe’s Magnetism \citep[POSSUM][]{Gaensler2025} 35 polarized sources per square degree over 20,630\,deg$^2$. To use these catalogs to study extragalactic magnetic fields, these data need to be complemented by optical information on spectroscopic redshift \citep[see, e.g.,][]{Vacca2015}. 
To date, such information is available for only a fraction of sources, for example about 10\,percent for the NVSS data \citep{Hammond2012} and 50\,percent for the LOw-Frequency ARray (LOFAR) Two-metre Sky Survey (LoTSS) data \citep{OSullivan2023}. 

The Faraday depth contains information about the physical properties of all magneto-ionic media along the line of sight, such as the Milky Way, galaxy clusters, and filaments.  
To isolate information about extragalactic magnetic fields, an accurate and reliable reconstruction of the Galactic Faraday rotation is required. 
\cite{Oppermann2015} showed that, due to observational uncertainty, the extragalactic Faraday rotation cannot be calculated simply as the difference between the observed Faraday rotation and the Galactic foreground. Assuming that the Galactic Faraday rotation is correlated and that the extragalactic Faraday rotation and observational noise are completely uncorrelated in the sky, they provided for the first time a simultaneous Bayesian reconstruction of the Galactic and extragalactic Faraday rotation, finding that the extragalactic contribution is well described by a zero-mean Gaussian with a standard deviation of approximately 7\,rad/m$^2$ \citep[see also, e.g.][]{Schnitzeler2010}. Building on this work, \cite{Vacca2015,Vacca2016} developed a Bayesian algorithm that relies on Gibbs sampling to disentangle and further  characterize the multiple contributions to the extragalactic Faraday rotation. The algorithm follows a two-step process: a sample for
the extragalactic contribution $\phi_{\rm e}$ is drawn from its posterior and then, holding it fixed, a new sample of the parameters $\Theta$ of the extragalactic Faraday rotation model is drawn from the conditional probability \textit{P}($\Theta | \phi_{\rm e}$). Throughout the entire process, the Galactic power spectrum, the Galactic profile, and the observed noise variance correction factors are held fixed at the values published by \cite{Oppermann2015}, see \cite{Vacca2016} for more details.

Recently, \cite{Hutschenreuter2022}, significantly improved
the reconstruction of the Galactic Faraday rotation sky, processing nearly all available Faraday depth datasets. This work represents an improvement over the previous one due to the larger data set and to the algorithmic technical advances based on the use of a new correlation structure model presented in \cite{Arras2022} and the use of Metric Gaussian Variational Inference \citep[MGVI, ][]{Knollmuller2019}.
In this paper, we extend \cite{Hutschenreuter2022}'s algorithm to statistically detect, distinguish, and characterize an excess in Faraday rotation variance intrinsic to the emitting radio source from that associated with the large-scale structure of the Universe.
This represents a significant improvement over the approach presented in \cite{Vacca2016} and other previous works, since, for the first time, this algorithm performs a simultaneous separation of various extragalactic components from the Galactic and noise contribution to the Faraday rotation. This approach allows for a better separation of the different contributions to the Faraday rotation and thus a more reliable estimate of the extragalactic Faraday effect. This represents a step forward towards an accurate separation of Galactic small-scale structures and extragalactic contributions, which requires an extremely dense RM-grid obtained over a wide frequency range.

Distinguishing between Galactic, extragalactic, and noise contributions is crucial to studying large-scale extragalactic magnetic fields and, thus, the history of cosmic magnetism. The magnetization of galaxy clusters is best studied through centimeter-wavelength observations and allows us to trace the evolution of magnetic fields over cosmic time. Along filaments and voids of the cosmic web, magnetization is less influenced by structure-formation processes and
can therefore be used to directly constrain magnetogenesis scenarios. Recently, \cite{Carretti2022} demonstrated that meter-wavelength observations can provide a window into the magnetization of low-density, weakly magnetized environments, such as filaments of the cosmic web. Using the LoTSS RM catalog \citep{OSullivan2023}, \cite{Carretti2023,Carretti2025} conclude that Faraday rotation at LOFAR frequencies is dominated by the contribution of cosmic filaments and that primordial magnetogenesis scenarios are favored. According to their results, the magnetic field has an average strength between 11 and 15\,nG at \textit{z} = 0 and evolves with redshift as $(1+\textit{z})^{\alpha}$ with $\alpha=2.3-2.6$.

In this manuscript, we present our algorithm and demonstrate its performance using synthetic data. The paper is organized as follows. In Sect.\,\ref{methodology}, we describe the methodology underlying our approach and the modeling adopted in this work. In Sect.\,\ref{syntheticcat}, we summarize how we construct synthetic catalogs for testing purposes. In Sect.\,\ref{results} and in Sect.\,\ref{discussion}, we present and discuss our results. Finally, in Sect.\,\ref{conclusions}, we present our summary and conclusions. 
Expectations for future radio polarimetric surveys obtained with the Square Kilometre Array (SKA) telescopes were presented in  \cite{Vacca2026}. The application of the algorithm to existing radio-optical catalogs is currently in progress and will be covered in a separate manuscript. 
In the following, we use a $\Lambda$CDM cosmology with $\textit{H}_0 =67.3$\,km/s/Mpc, $\Omega_0$ = 0.315, and $\Omega_{\Lambda}$ = 0.685.

\section{Methodology}
\label{methodology}

The inference algorithm presented in this article is based on a Bayesian approach within the framework of Information Field Theory \citep[IFT, ][]{Ensslin2009}. The algorithm is implemented in version 8 of the Python package NIFTy\footnote{\protect{\url{ https://ift.pages.mpcdf.de/nifty/}}} \citep{Arras2019}, a  library for Bayesian signal inference
based on IFT, particularly suited to tackling problems dominated by noisy data and characterized by high dimensionality. Below, we  discuss the a priori assumptions and the forward likelihood model.
A sketch of the model is presented in Appendix\,\ref{appendixA}.

Our data \textit{d} is a vector containing the Faraday rotation in the direction of $\textit{N}_{\rm los}$ point-like radio background sources, i.e.,  $\textit{d} = \{\textit{d}_i\}_{i=1}^{\textit{N}_{\rm los}} = \textit{d}_1, \textit{d}_2, ..., \textit{d}_{\rm \textit{N}_{\rm los}}$. Because different lines of sight to the sources probe different combinations of Faraday rotating environments, these Faraday depth measurements provide information about the magnetic fields in our Galaxy and in extragalactic environments. 
\cite{Hutschenreuter2022} model the observed Faraday rotation for a set of extragalactic radio sources as a Galactic Faraday rotation plus a noise term,
\begin{equation}
    \textit{d}_{\rm i}=\phi_{\rm gal, i}+\tilde{\textit{n}}_{\rm i},
    \label{galonly}
\end{equation}
where the noise term $\tilde{\textit{n}}_{\rm i}$ contains the known observational uncertainty and any extragalactic component or unresolvable small-scale (even Galactic) structure. The noise term, including the extragalactic Faraday rotation, is assumed to have a Gaussian distribution. The variance of this Gaussian is proportional to the measurement variance $\sigma^2_{\rm n}$, increased by a factor $\eta$,
\begin{equation}
\tilde{\textit{N}} = \rm diag~(\eta \sigma_{n}^2)
\label{etafactor}
\end{equation}
where the factor $\eta$ is inferred from the data, differs for each line of sight, and takes into account possible corrections for unreliable estimates of the observational noise and the extragalactic contribution.

In this work, for lines of sight with auxiliary information (e.g., optical spectroscopic redshifts), we further separate the Faraday rotation and model it as a sum of Galactic and extragalactic components plus observing noise,
\begin{equation}
    \textit{d}_{\rm i}=\phi_{\rm gal, i}+\phi_{\rm eg, i} + \textit{n}_{\rm i}.
    \label{galplusegal}
\end{equation}
In this new version of the algorithm, a revised stochastic generative model is set up for the data. This model includes physical quantities of interest, such as the dispersion of the Faraday rotation expected per unit-length of the overall large-scale extragalactic structure between the source and the observer, as model components (further disentanglement is also possible, considering different cosmic environments such as galaxy cluster, filament, etc.). The parameters of the generative model are then inferred conditional to the observed data. The algorithm produces a Bayesian posterior distribution of all model parameters. Field-like quantities, such as the Galactic Faraday sky, must also be inferred. For this reason, the inference problem is very high-dimensional. To address this problem in finite computational time, approximate variational inference methods, namely MGVI, and geometrical Variational Inference \citep[geoVI][]{Frank2021}, are used, which allow for a more accurate and stable posterior approximation. Since MGVI is generally faster, we start the inference process with it and subsequently refine the result with geoVI. geoVI is useful for highly nonlinear posterior distributions that deviate significantly from a Gaussian. This algorithm maps the posterior distribution from the space where the model parameters are mapped to a space where the posterior distribution can be approximated by a Gaussian. Specifically, by minimizing the Kullback-Leibler divergence, geoVI finds the optimal location where the posterior distribution can be approximated by a normal distribution. This approach has already been successfully applied in other works, e.g. \cite{Hutschenreuter2024} and \cite{Arras2022}. We stress that the technical advances consisting in the simultaneous separation of Galactic and extragalactic Faraday rotation and the use of variational inference represent a significant improvement over the previous version of the algorithm described in \cite{Vacca2016}. 

Overall, our approach relies on the availability of two catalogs: 
\begin{itemize}
 \setlength{\itemsep}{5pt}

    \item[-]
one catalog contains the information about the Faraday rotation and its uncertainty that can only be used for the inference of the Galactic contribution (Eq.\,\ref{galonly});

\item[-] a second catalog also contains auxiliary information (such as redshift, total intensity flux density, etc., see below) that can be used to simultaneously constrain the Galactic and extragalactic contributions (Eq.\,\ref{galplusegal}). Auxiliary information is typically available for only a fraction of the sources for which Faraday depth data are provided.
\end{itemize}

\subsection{Galactic Model}

We model the Faraday rotation of our Galaxy according to \cite{Hutschenreuter2020},
\begin{equation}
    \phi_{\rm gal}=\textit{e}^{\rho}\chi,
    \label{galcontr}
\end{equation}
where $\rho$ and $\chi$ are two Gaussian fields on the sky with an unknown correlation structure that is inferred by the algorithm. Assuming no correlation between the thermal gas density and the parallel component of the magnetic field in our own Galaxy, $e^{\rho}$ mimics the dispersion measure, while $\chi$ mimics the average of the magnetic field. Furthermore, \cite{Hutschenreuter2024}
implemented a model that includes dispersion and emission measure as observables to discriminate between (i) the line-of-sight averaged magnetic field component, (ii) the electron density-weighted magnetic field component and (iii) the dispersion measure over the entire sky. Unlike that work, we do not resort to observed dispersion and emission measure, but rely only on Faraday depth measurements, adopting the model presented in \cite{Hutschenreuter2020} and \cite{Hutschenreuter2022}. A detailed description of the Galactic and noise inference is provided in these works, as well as in \cite{Oppermann2012}.

\subsection{Extragalactic Model}
\label{sec_egmodel}

We assume that the extragalactic Faraday rotation can be
approximated by a zero-mean Gaussian, 
\begin{equation}
\langle \phi_{\rm e,i}\rangle = 0, 
\end{equation}
with variance
\begin{equation}
\begin{split}
\langle \phi_{\rm e,i}^2\rangle \approx \textit{a}_0^2\int_0^{\textit{z}_{\rm i}}\langle \textit{n}_0^2\rangle \langle \textit{B}_{\rm l_0}^2\rangle\Lambda_{\rm l_0}(1+\textit{z})^4\frac{\textit{c}}{\textit{H(z)}}d\textit{z},
\end{split}
\label{RMvar}
\end{equation}
where \textit{z} is the redshift of the source, \textit{a}$_0$ is a normalization constant, \textit{c} is the speed of light, \textit{H(z)} is the Hubble parameter, and $\langle \textit{n}_0\rangle$, $\langle \textit{B}_{\rm l_0}\rangle$, and $\Lambda_{\rm l_0}$ are the present-day thermal gas density, magnetic field, and magnetic field length scale\footnote{We define the length scale as
\begin{equation} \notag
\Lambda_{\textit{l}_0}:= \int \mathrm{d}\textit{l}_0^{\prime}\frac{\langle \textit{B}(\textit{l}_0)\textit{B}(\textit{l}_0^{\prime})\rangle}{\langle \textit{B}(\textit{l}_0)^2\rangle}.
\end{equation}}, as derived by \cite{Vacca2016}. 
From an observational point of view \citep[see, for example,][]{Clarke2004,Schnitzeler2010,Oppermann2015}, we have no reason/evidence for a positive/negative mean in the extragalactic Faraday rotation distribution and, therefore, we assume a mean of zero. However, from a theoretical point of view, we note that we can expect a non-zero mean as, for example, in case of primordial magnetic fields, generated during either phase transitions or inflation, see \cite{Durrer2013} for a review.

In general, the extragalactic Faraday rotation variance in Eq.\,\ref{RMvar} can be expressed as a function of a vector of parameters $\Theta$, 
\begin{equation}
\langle \phi_{\rm eg, i}^2 \rangle = \sigma_{\rm eg, i}^2(\Theta).
\label{egmodel}
\end{equation}
$\Theta$ is an \textit{N}-dimensional vector, where \textit{N} is the number of parameters used to represent the extragalactic Faraday depth variance and its elements are the parameters we wish to infer. 
Following \cite{Vacca2016}, we assume that the extragalactic Faraday rotation variance consists of two terms, 
\begin{equation}
\sigma_{\rm eg, i}^2(\Theta)=\sigma_{\rm int, i}^2(\Theta)+\sigma_{\rm env, i}^2(\Theta),
\end{equation}
a Faraday rotation variance dependent on the luminosity of the emitting radio source, 
\begin{equation}
\sigma_{\rm int, i}^2(\Theta) 
=\left(\frac{\textit{L}_{i}}{\textit{L}_0}\right)^{\chi_{\rm lum}}\frac{e^{\chi_{\rm int, 0}}}{(1+\textit{z}_{i})^4},
\label{sigmaint}
\end{equation}
and a redshift-dependent Faraday rotation associated with the medium between the source and our Galaxy,
\begin{equation}
\sigma_{\rm env, i}^2(\Theta)
= \textit{e}^{\chi_{\rm env, 0}}\frac{1}{\textit{D}_0}\int_0^{\textit{z}}\frac{\textit{c}}{\textit{H(z)}}(1+\textit{z}_{i})^{4+\chi_{\rm red}}d\textit{z},
\label{sigmaenv}
\end{equation}
where \textit{L}$_{i}$ is the luminosity of the source, and \textit{L}$_0=10^{27}$\,W/Hz and \textit{D}$_0=5000$\,Mpc are normalization constants. This modeling allows us to explore possible dependence of the extragalactic Faraday rotation on the total intensity luminosity of the source. After subtracting the Faraday rotation associated with our Galaxy in NVSS data, \cite{Pshirkov2015} find that the residual RM positively correlates with the luminosity in total intensity of the sources at 1.4\,GHz, with higher luminosity sources showing higher residual RM. If this dependence is not properly modeled, they show that an apparent evolution of rotation measure with redshift can be observed likely caused by different intrinsic properties of brighter high-redshift sources. However, different and possibly more appropriate models can be adopted in future work.

Overall, the parameters to be inferred are $\Theta = \{\chi_{\rm int, 0},~ \chi_{\rm env, 0},~ \chi_{\rm lum},~ \chi_{\rm red}\}$, where $\chi_{\rm lum}$ is the luminosity dependence, $\chi_{\rm red}$ is the redshift dependence, and $\sigma_{\rm int,  0}^2=\textit{e}^{\chi_{\rm int, 0}}$ and $\sigma_{\rm env, 0}^2=\textit{e}^{\chi_{\rm env, 0}}$ represent the pure intrinsic and environmental Faraday rotations, respectively, where the environmental contribution is intended per unit length. Pure means that all redshift and luminosity dependencies are absorbed by the multiplicative factors, mainly through $\chi_{\rm lum}$ and $\chi_{\rm red}$. These parameters characterize the Faraday rotation intrinsic to the emitting radio sources and that associated with extragalactic environments along the line of sight (e.g., galaxy clusters, filaments, voids, sheets) and their dependence on physical quantities such as the redshift and luminosity of the emitting sources. Since the Universe is uniform and isotropic, and we expect all the sources to belong to the same physical ensemble, these parameters are assumed to be the same for each line of sight.

The overall variance of the extragalactic Faraday rotation is treated here statistically as a systematic noise variance. The algorithm separates the different contributions to the Faraday rotation relying on the fact that the Galactic component is characterized by angular correlations on the sky, the extragalactic and noise terms instead are not.
Consequently, uncorrelated small-scale structures due to the Galactic magneto-ionic medium could be absorbed by the noise or by the extragalactic term.
Conversely, correlated extragalactic structures end up in the Galactic Faraday rotation. The extragalactic Faraday contribution inferred by our algorithm includes only the extragalactic Faraday rotation resulting from the uncorrelated component of the magnetic field.
An approach that exploits the angular correlation of the extra-galactic Faraday rotation will be addressed in future work.

\section{Synthetic catalogs}
\label{syntheticcat}
We tested the inference setup with the help of synthetic catalogs generated on the basis of observed data. In particular, we considered the Faraday depth  of \cite{VanEck2023}, version number 1.2.0, complemented by the spectroscopic redshifts provided by \cite{OSullivan2023} for the LoTSS data points and by \cite{Hammond2012} for the NVSS data points. According to our modeling, we exploit sources for which both radio and optical information is available to constrain the Galactic and extragalactic contributions, while we use sources for which only radio information is available to infer the Galactic Faraday rotation using exclusively the \cite{Hutschenreuter2022} model.
For simplicity, in generating the synthetic-catalogs, we kept the source coordinates identical to those in the observed Faraday depth catalogs but modified the Faraday rotation, its uncertainty, the redshift, and the total intensity luminosity. We assumed that the uncertainty in redshift, luminosity, and coordinates was negligible.
Below, we describe how the Galactic, extragalactic, and noise contributions were generated for the sources in these catalogs.

\subsubsection*{Galactic contribution}
\label{galcon}
\begin{figure*}
\centering
\includegraphics[width=0.33\linewidth]{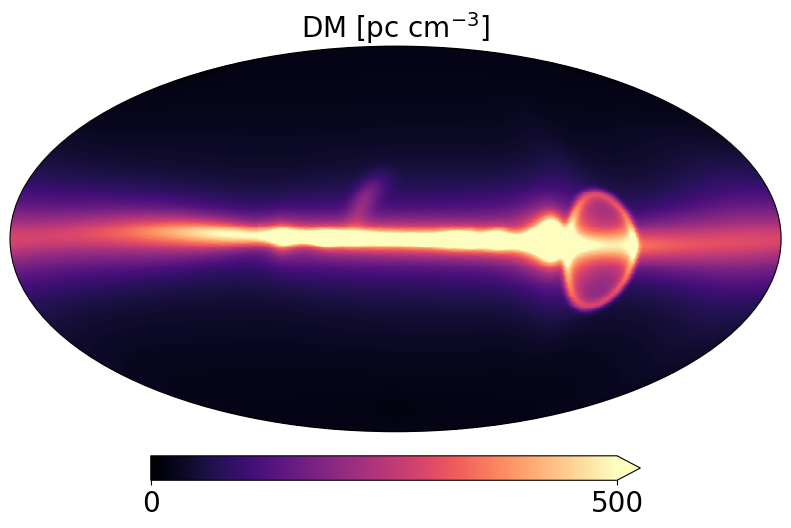}
\includegraphics[width=0.33\linewidth]{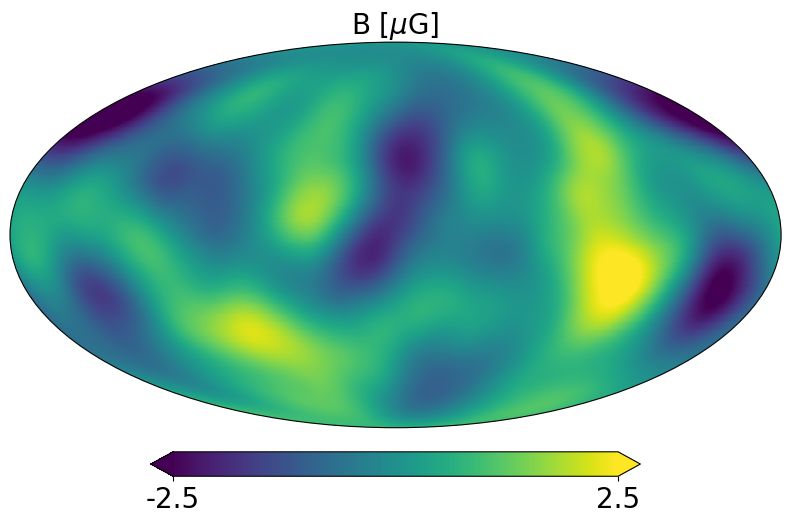}
\includegraphics[width=0.33\linewidth]{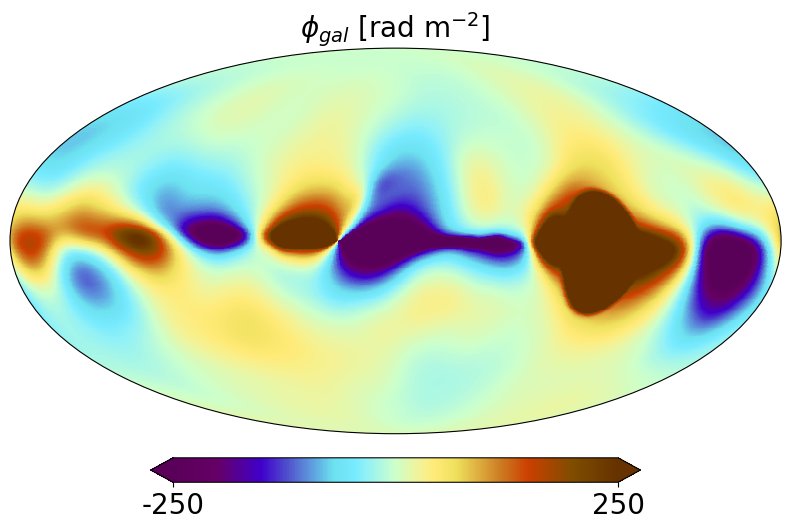}\\
\includegraphics[width=0.33\linewidth]{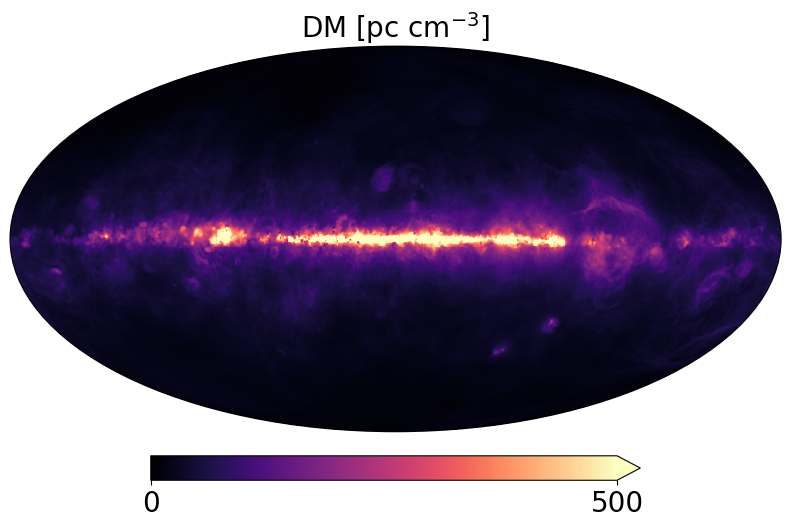}
\includegraphics[width=0.33\linewidth]{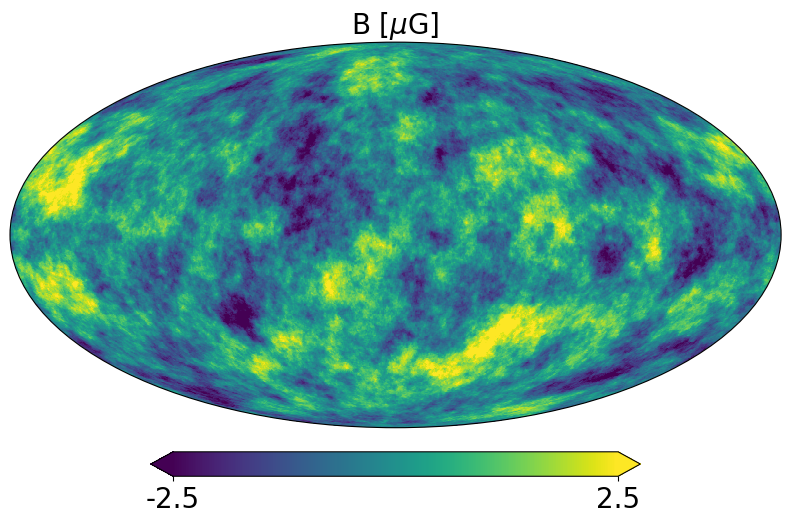}
\includegraphics[width=0.33\linewidth]{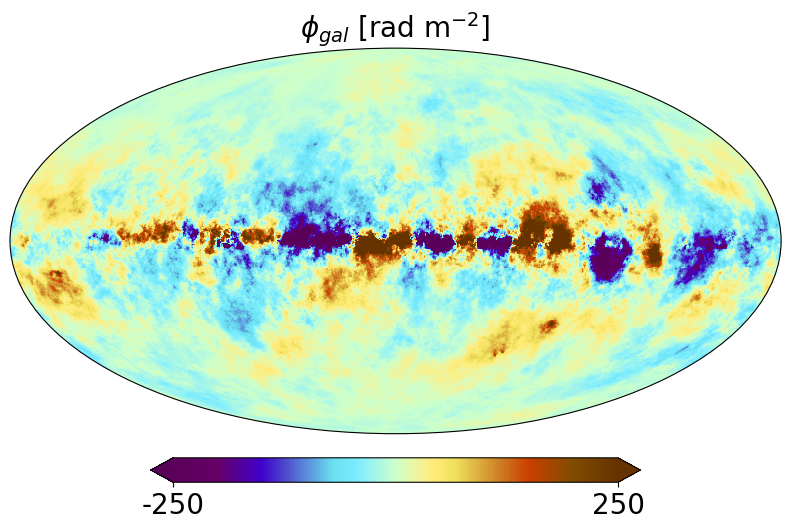}\\
\caption{Left panels: free electron distribution model by \cite{Yao2017} on the top and by \cite{Hutschenreuter2024} on the  bottom. Middle panels: synthetic component of the Galactic magnetic field along the line of sight corresponding to a slope of the magnetic field power spectrum of $\gamma=-8$ on the top and of $\gamma=-3$ on the bottom. Right panels: synthetic Galactic Faraday rotation corresponding to a slope $\gamma=-8$ on the top and $\gamma=-3$ on the bottom. Further details on Galactic magnetic field and Faraday rotation generation are available in Sect.\,\ref{galcon}.}
\label{fig1}
\end{figure*}

To calculate the Galactic contribution, we computed the Faraday rotation as
\begin{equation}
    0.81 \times DM \times \textit{B}_{\rm l}~~~{\rm rad/m^2}
    \label{galeq}
\end{equation}
where DM is the dispersion measure and $\textit{B}_{\rm l}$ is the average magnetic field along the line of sight. In tests to evaluate the algorithm's performance, we adopt as a dispersion measure either the smooth model for the free electron distribution of \cite{Yao2017} see Sect.\,\ref{functioning} and Sect.\,\ref{slope},  
shown in the top left panel of Figure\,\ref{fig1}, or the mean of the reconstruction of \cite{Hutschenreuter2024} see Sect.\,\ref{etafactor},  
shown in the bottom left panel of Figure\,\ref{fig1}. For the magnetic field, we draw a random sample generated considering a power spectrum $|\textit{B}_{\ell}|^2$ that varies as a function of the wave number $\ell$ as
\begin{equation}
|\textit{B}_{\ell}|^2=\frac{\textit{P}_0}{1.0 + \left( \frac{\ell}{\ell_0} \right) ^{-\gamma}},
\label{b_ps}
\end{equation}
where \textit{P}$_0$ is the normalization of the power spectrum, $\ell_0$ is the characteristic wave number, and $\gamma$ is the slope of the power spectrum. 
This results in a random Gaussian component of the magnetic field along the line of sight. 
In this work, we adopted $\ell_0=5.0$, and $\gamma$ ranging from -2 to -8. The normalization was chosen to have a magnetic field with variance of one. In the middle panels of Figure\,\ref{fig1}, we show, as an example, synthetic Galactic magnetic field distributions generated as described above, with a power spectrum slope of $\gamma=-8$ and $\gamma=-3$, respectively. The right panels of the same figure show the corresponding synthetic Galactic Faraday rotation, each obtained by combining the dispersion measure and the magnetic field distributions shown in the left and middle panels in the corresponding row. The magnetic field and Galactic Faraday rotation images obtained using a slope
$\gamma=-3$ include small-scale fluctuations, instead for 
$\gamma=-8$ the fluctuations are mainly on larger scales. The different fluctuation scales are due primarily to the different slopes of the magnetic field power spectrum, but also in part to the different scales present in the dispersion measure images. 

\subsubsection*{Extragalactic contribution}
\begin{figure}
\centering
\includegraphics[width=9 cm]{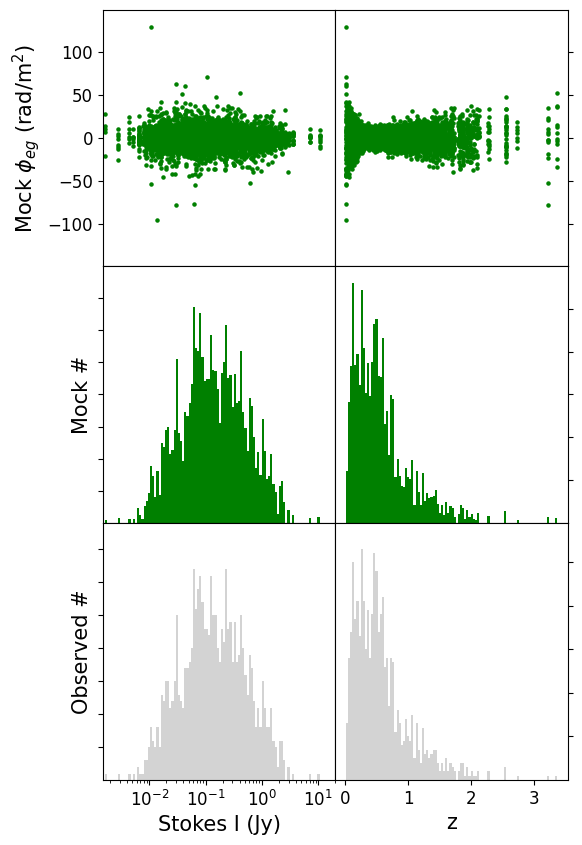}
\caption{Top panels: in green, synthetic $\phi_{\rm eg}$ with respect to Stokes I (left) and redshift (right), corresponding to an overall extragalactic Faraday rotation of $\approx$\,8\,rad/m$^2$. Middle panels: in green, synthetic Stokes I distribution (left) and redshift distribution (right). Bottom panels: in light gray, observed Stokes I  distribution for sources with redshift (left) and observed redshift distribution (right) as reported in \cite{OSullivan2023}.}
\label{fig2}
\end{figure}

In order to disentangle the extragalactic contribution from the Galactic and noise term, our algorithm requires accurate redshift measurements associated with the observed radio sources. Indeed, this complementary information allows us to compute the distance covered by the signal in its way to the observer. However, as described in Sect.\,\ref{intro}, in present-day radio catalogs, such measurements are available only for a limited number of sources. Therefore, in the following, we include an extragalactic term only for a fraction of sources.
The extragalactic contribution was randomly drawn from a Gaussian distribution with zero mean and variance calculated using the model described in Sect.\,\ref{sec_egmodel}. Synthetic redshifts and luminosities were extracted independently from the corresponding observed distributions, similarly to what was done in \cite{Vacca2016}. Most catalogs available in the literature provide Stokes I measurements rather than luminosities. These values represent either intensities or flux densities. For the purposes of our study, we considered them as flux densities and calculated luminosities according to $\textit{L}=4\pi \textit{D}_{\rm L}^2\textit{F}$, where $\textit{D}_{\rm L}$ is the luminosity distance and \textit{F} the flux density. In Figure\,\ref{fig2} we show as an example the synthetic contributions of the extragalactic Faraday rotation with respect to Stokes I and redshift, as well as the observed and synthetic distributions of Stokes I and redshift. 

\subsubsection*{Faraday rotation uncertainty}
Finally, a noise term has been added. To this end, 
we produce synthetic uncertainties of the Faraday rotation by 
sampling directly from observed uncertainty distributions.
For each line of sight, we produce noise contributions by randomly drawing from a Gaussian distribution with zero mean and variance given by these synthetic uncertainties.

To evaluate the code's performance, we first considered a very small Faraday rotation uncertainty for all the sources in the catalog, by sampling the LoTSS uncertainty distribution, characterized by a median value of approximately 0.06\,rad/m$^2$. 
\begin{figure}
\centering
\includegraphics[width=7 cm]{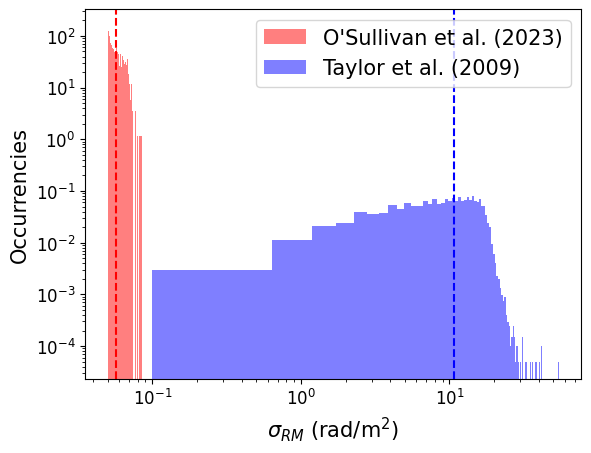}
\caption{Histogram of the uncertainty in Faraday depth measurements from the LoTSS catalog \citep{OSullivan2023} in red and from the NVSS catalog \citep{Taylor2009} in blue. The vertical dashed lines indicate the median of each distribution.}
\label{fig3}
\end{figure}
Furthermore, we tested the algorithm with catalogs including measurements reflecting both NVSS and LoTSS uncertainties, hereafter observed-like catalogs. The distribution of uncertainties in the NVSS and LoTSS catalogs is shown in Figure\,\ref{fig3}.

\section{Results}
\label{results}
Following the above prescriptions, we produced synthetic catalogs assuming different Galactic magnetic field power spectra, different overall extragalactic contributions, different cuts in Galactic latitude of the observed radio sources and different uncertainties in Faraday rotation, reflecting currently publicly available Faraday measurements. 
Overall, we performed tests considering the following synthetic catalogs: 
\begin{enumerate}
\item Faraday rotation catalogs with a Galactic Faraday rotation on large spatial scales, a smooth distribution of free electrons, an overall extragalactic standard deviation of approximately 5-10\,rad/m$^2$, and reliable, small uncertainty in the Faraday rotation. The results of these tests are presented in Sect.\ref{functioning};
\item Faraday rotation catalogs with Galactic Faraday rotation on a range of spatial scales (from large to small), a smooth distribution of free electrons, an overall extragalactic standard deviation of approximately 5-10\,rad/m$^2$, and reliable, small uncertainty in Faraday rotation. The results of these tests are presented in Sect.\ref{slope}; 
\item NVSS-like and LoTSS-like catalogs, with an overall extragalactic contribution of $\approx$10\,rad/m$^2$ \citep[see, e.g., ][]{Oppermann2015} and $\approx$1.5\,rad/m$^2$ \citep[see, e.g., ][]{Carretti2022}, respectively, and Faraday uncertainties consistent with those of NVSS and LoTSS. We considered magnetic field power spectra and distribution of free electrons in agreement with the observations and extragalactic radio sources located above an absolute Galactic latitude $b$ of 0, 20, and 45\,degrees. The results of these tests are presented in Sect.\ref{etafactor};
\item Faraday rotation catalogs with a Galactic Faraday rotation on large spatial scales, a smooth distribution of free electrons, no extragalactic contribution, and reliable, and small uncertainty in the Faraday rotation. For these catalogs inference has been performed assuming a one component-model, in order to test for false positive detections. The results of these tests are presented in Sect.\ref{no_eg_contr}.
\end{enumerate}

We assume Gaussian priors on all the extragalactic parameters to infer. We opted for a zero mean and a standard deviation equal to or larger than 0.5, to ensure a non-negligible probability in correspondence of the synthetic value of the parameter of interest. Due to computational constraints, the spatial resolution used during inference is approximately 0.05\,deg$^{2}$.

\begin{figure*}
\centering
\includegraphics[width=0.7\linewidth]{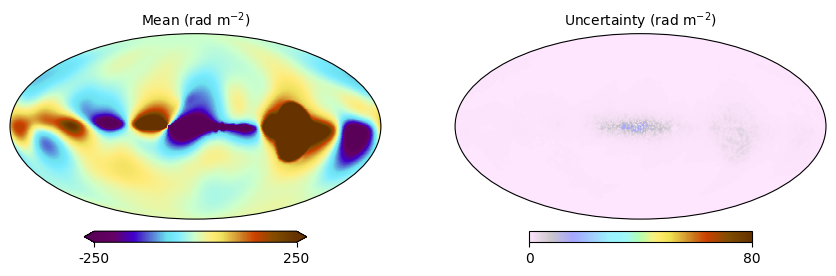}\\
\vspace{0.5cm}
\includegraphics[width=0.7\linewidth]{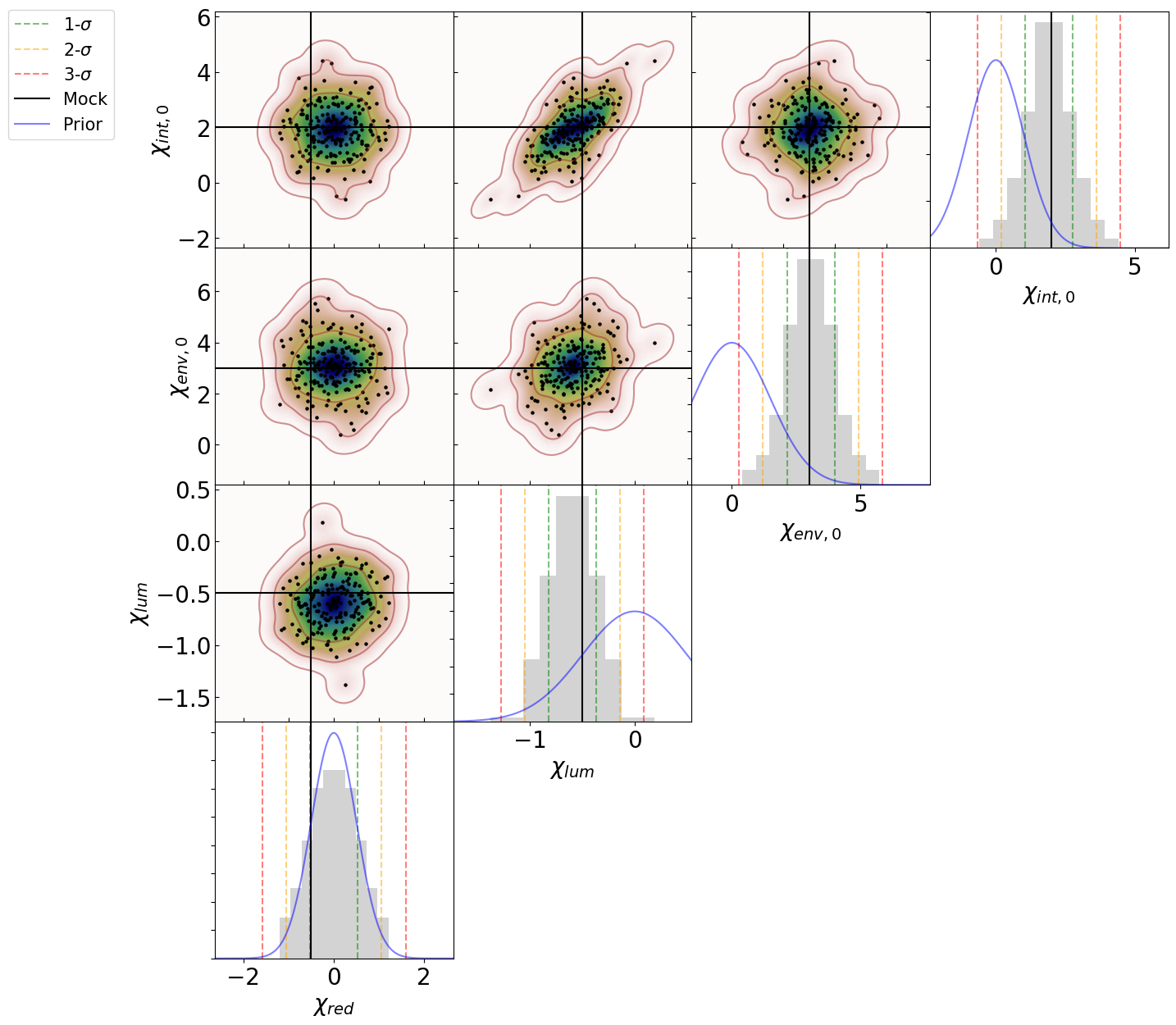}
\caption{Results for the configuration described in Sect.\,\ref{functioning}. Top panels: mean (left) and uncertainty (right) of the posterior distribution of the Galactic Faraday sky. Bottom panels: 2-dimensional marginalization (in colors) and 1-dimensional marginalization (gray histograms) of the posterior distribution. The black line represents the synthetic value of the parameter. Brown contours represent the 1-, 2-, and 3-$\sigma$  levels of the 2-dimensional marginalizations of the posterior distribution. The green, orange, and red dashed lines represent the 1-, 2-, and 3-$\sigma$ levels of the 1-dimensional marginalization of the posterior distribution. Blue lines show the prior distributions. Black dots are randomly drawn samples from the posterior distribution. Mirror images of the samples around the mean have also been included.
}
\label{fig4}
\end{figure*}

\subsection{Large-scale Galactic power spectrum}
\label{functioning}

We first perform a test assuming
\begin{itemize}
    \item a magnetic field power spectrum for the Galactic term on large-scale, i.e. $\gamma=-8.0$, see Eq.\,\ref{b_ps};
    \item the smooth model for the distribution of free electrons by \cite{Yao2017} as dispersion measure;
    \item an overall standard deviation in the extragalactic contribution of $\approx$5-10\,rad/m$^2$;
    \item redshift, Stokes I, and a reliable, small uncertainty in the Faraday rotation, sampled from the observed LoTSS distribution \citep{OSullivan2023}.
\end{itemize}
Due to the large-scale magnetic field power spectrum and to the smooth spatial distribution of the dispersion measure, the synthetic Galactic Faraday rotation does not contain any small-scale features, so the small scale contribution can be entirely attributed either to the extragalactic Faraday rotation or to the noise. 
The synthetic catalog includes approximately sixty thousand sources, comparable to the number of Faraday depth measurements presented in the \cite{VanEck2023} catalog used in this work (version number 1.2.0). All sources were used to infer the Galactic Faraday sky, while only about 15-20\,percent of the sources (about ten thousand sources) were used to infer the extragalactic model parameters. These sources were selected to have an absolute value of the Galactic latitude greater than 45.0\,degrees to ensure low Galactic contamination.  Only for these sources an extragalactic term was included in the synthetic catalog. For testing purposes, the Faraday rotation of the remaining sources includes only a Galactic contribution. However, we note that if an extragalactic contribution was also present for these sources, it would be absorbed by the $\eta$ factors, see \cite{Hutschenreuter2022}.

To better evaluate the algorithm's performance, we analyze the inference results for Galactic and extragalactic Faraday rotation.
In the top panels of Figure\,\ref{fig4} we show the results for the Galactic Faraday sky. Compared to the top-right panel in Figure\,\ref{fig1}, the reconstruction shows a high level of fidelity, with an uncertainty typically of a few rad/m$^2$ and up to 20-30\,rad/m$^2$ at the Galactic center alone. The presence of a Galactic foreground that is not perfectly known imprints on the RM dispersion, mostly in creating regions of enhanced residual dispersion in the Galactic disk, and to a lesser degree at higher latitudes. In the bottom panel of Figure\,\ref{fig4}, we show the results for the extragalactic Faraday rotation. We show in colors the two-dimensional marginalizations of the posterior distribution for each pair of parameters. Brown contours indicate the 1-, 2-, and 3-$\sigma$ level of the 2-dimensional marginalized posterior distribution. Black lines show the synthetic values of the parameters used in generating the synthetic catalog. Gray histograms represent the 1-dimensional marginalizations of the posterior distributions; dashed green, orange, and red lines indicate their 1-, 2-, and 3-$\sigma$ levels. Blue lines show the prior distribution adopted for each parameter. The synthetic value for each parameter is recovered within 1-$\sigma$, although some correlations can be identified, the most prominent of which is  between $\chi_{\rm int,0}$ and $\chi_{\rm lum}$. This correlation reflects a degeneracy between the parameters, see Eq.\,\ref{sigmaint}: a given value of $\sigma^2_{\rm int, i}$ can be obtained from different combinations of $\chi_{\rm lum}$ and $e^{\chi_{\rm int, 0}}$.

\begin{figure}[h!]
\centering
\includegraphics[width=\linewidth]{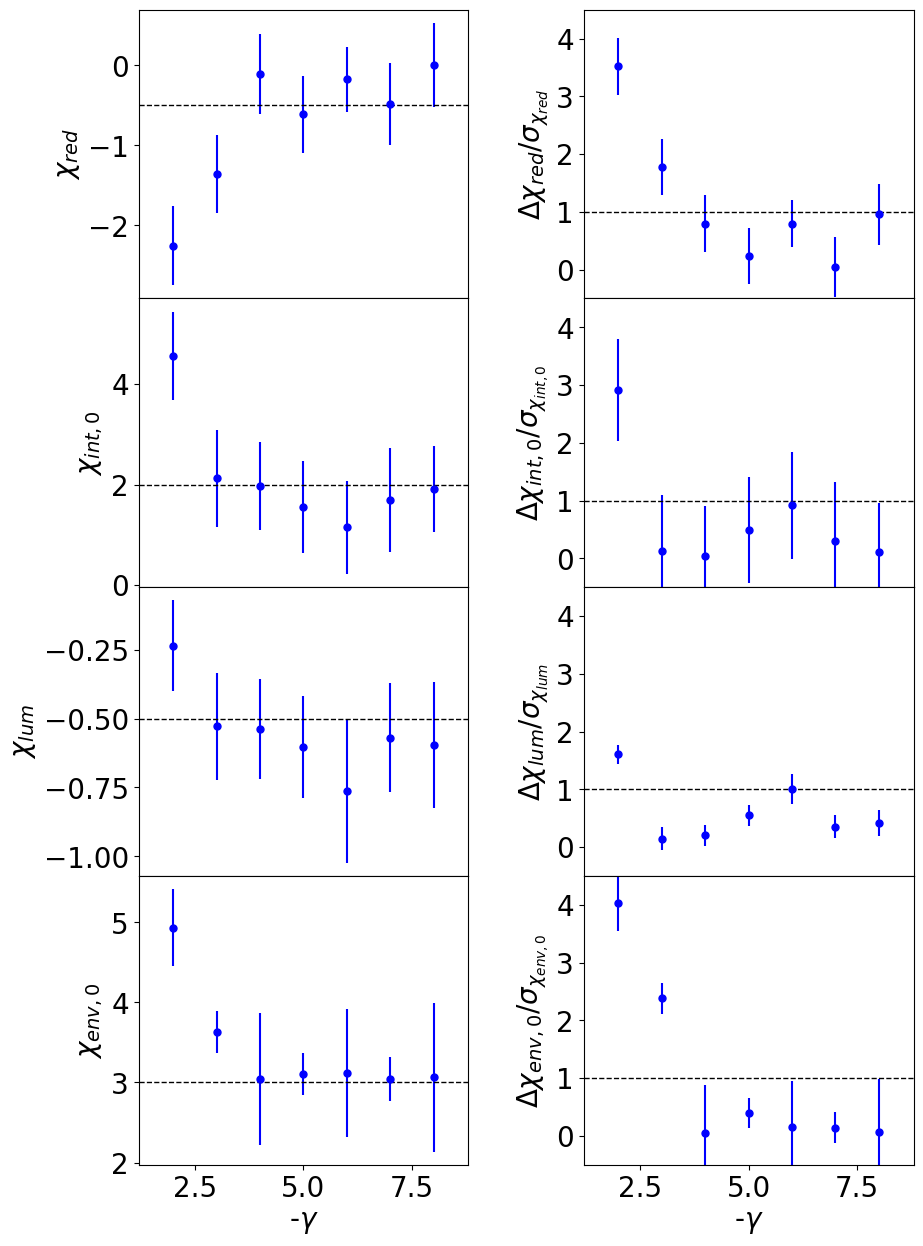}
\caption{Inferred values (left panels) and difference between the inferred value and the ground truth value in units of its a posteriori 1-$\sigma$ uncertainty (right panels) as a function of $-\gamma$, where $\gamma$ is the slope of the Galactic magnetic field power spectrum, for $\chi_{\rm red}$, $\chi_{\rm int,0}$, $\chi_{\rm lum}$, and $\chi_{\rm env, 0}$, from top to bottom respectively. The horizontal dashed line in each panel in the first column shows the ground truth value for each parameter, while in the second column a reference value of 1.}
\label{fig5}
\end{figure}

\subsection{Galactic power spectrum from large- to small-scales}
\label{slope}
To evaluate the algorithm's ability to correctly discriminate between Galactic and extragalactic Faraday rotation in presence of small-scale features in the Galactic term, we produced synthetic catalogs with the same configuration described in Sect.\,\ref{functioning} but with a slope of the Galactic magnetic field power spectrum $\gamma$ between -2 and -7, see Eq.\,\ref{b_ps}, in addition to the test with $\gamma=-8$ presented in the previous section. 
In Figure\,\ref{fig5} we show the results. For each extragalactic parameter, we show 
the inferred value and the corresponding uncertainty for different slopes of the Galactic magnetic field power spectrum. For comparison, we show the synthetic value of the parameter as a dashed line. 
For slopes $\gamma\leq -4$, performance is comparable, and synthetic parameter values can be recovered within 1-$\sigma$, meaning that for these slopes the algorithm can distinguish small-scale contributions of Galactic and extra-galactic origin, as well as the observing uncertainty, with high accuracy. Performance is also good for $\gamma=-3$. In this case, two parameters are recovered within 1-$\sigma$ and the other two within at most 2.5-$\sigma$. Using a slope of $\gamma=-2$, extragalactic parameter recovery is slightly worse, since the inferred values agree with the synthetic values between 1.5-$\sigma$ and 4-$\sigma$.

\begin{figure*}[h!]
\centering
\includegraphics[width=0.45\linewidth]{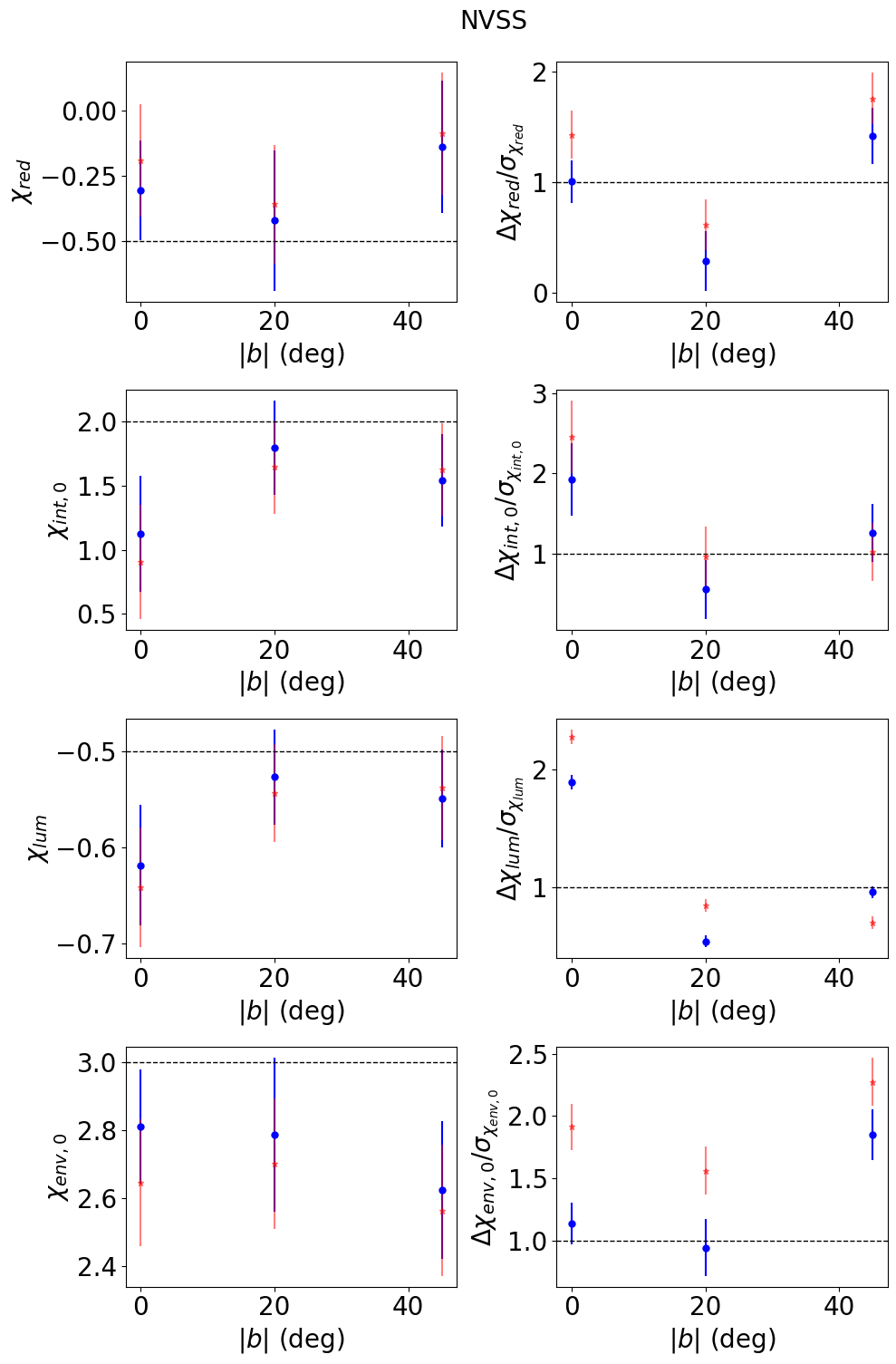}~~~\includegraphics[width=0.45\linewidth]{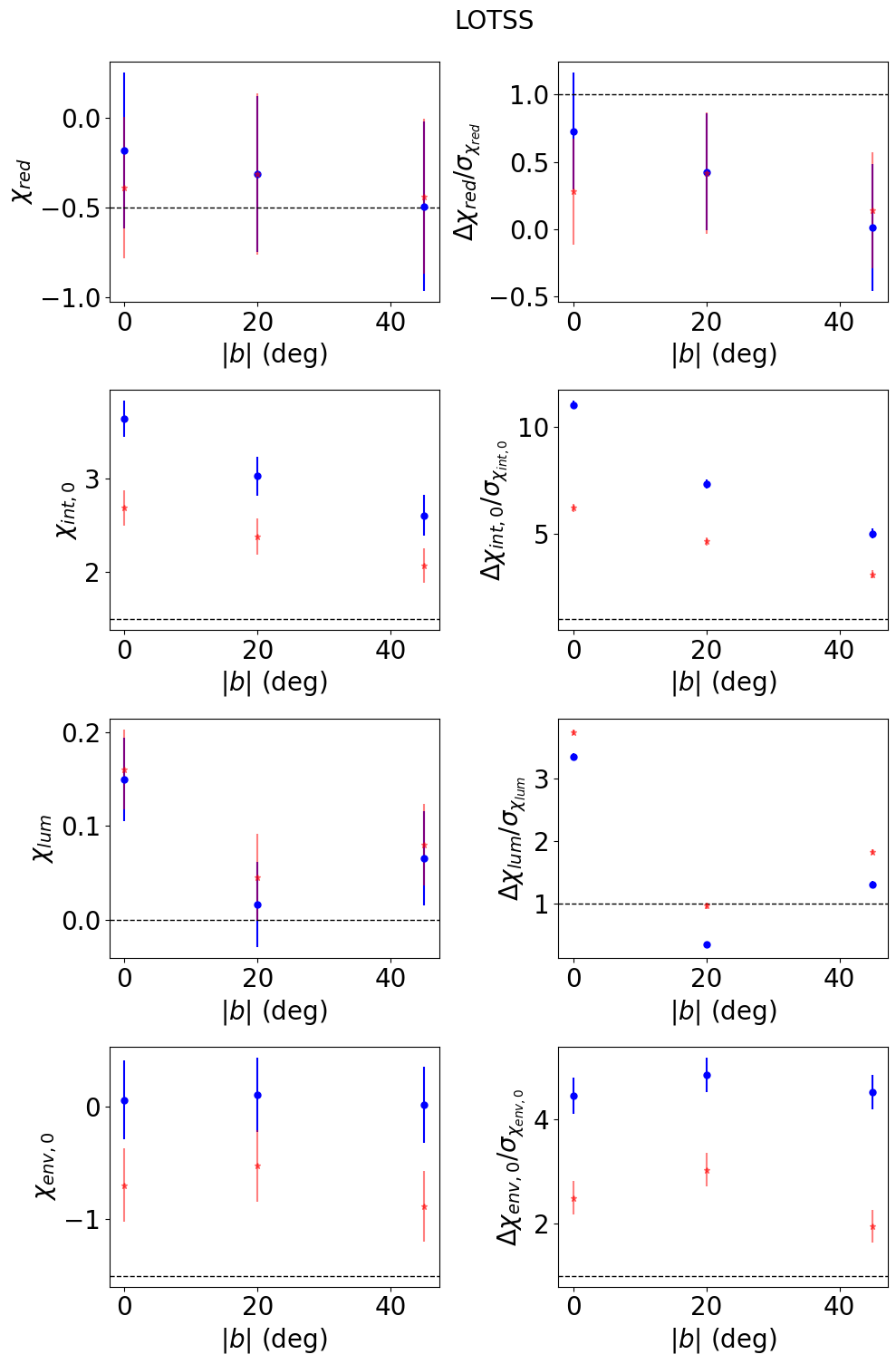}
\caption{Inferred values (first and third column) and difference between the inferred value and the ground truth value in units of its a posteriori uncertainty at 1-$\sigma$ (second and fourth column) for extragalactic sources populating the sky above the absolute Galactic latitude value shown in abscissa, for $\chi_{\rm red}$, $\chi_{\rm int,0}$, $\chi_{\rm lum}$, and $\chi_{\rm env, 0}$, from top to bottom respectively. Blue dots show results for a Galactic magnetic field power spectrum slope $\gamma=-3$, while red stars for  $\gamma=-4$. The horizontal dashed line in each panel in the first and third column shows the ground truth value for each parameter, while in the second and fourth column a reference value of 1.}
\label{fig6}
\end{figure*}

\subsection{Observed-like catalogs and Galactic latitude dependence}
\label{etafactor}

To evaluate the applicability of the algorithm to observed catalogs, we produced synthetic catalogs based on specifications of current observations. As described in Sect.\,\ref{syntheticcat}, we built observed-like catalogs whose Faraday rotation uncertainty reflects the NVSS and LoTSS uncertainties. Specifically, we distinguish among NVSS-like and LoTSS-like catalogs, that have been built as follows:
\begin{itemize}
\item[I] we consider a synthetic catalog with the same properties as the NVSS data. We sample the Faraday rotation uncertainty from the observed distribution for NVSS data points both for Galactic and extragalactic synthetic Faraday rotations. Based on the results of \cite{Hammond2012}, we assume that spectroscopic redshifts are available for about 4000 sources with no preferred Galactic latitude and adopt synthetic redshift values sampled from the observed distribution presented in that work. We assume an overall extragalactic Faraday rotation of about 10\,rad/m$^2$, comparable to that predicted using cm-wavelength Faraday depth measurements dominated by galaxy clusters. Sources distributed over all the sky have been considered, i.e. we did not apply any selection in Galactic latitude;
\item[II] we consider a synthetic catalog with the properties of the NVSS data for data points including solely the Galactic term and the LoTSS data for the points that also include a synthetic extragalactic Faraday rotation contribution. We sample the corresponding Faraday rotation uncertainty from the observed distributions for the NVSS and LoTSS data points, respectively. Based on the results of \cite{OSullivan2023}, we assume that spectroscopic redshifts are available for approximately 900 sources. These sources are assumed to have a Galactic latitude $|b|>20.0$\,degrees and an overall extragalactic Faraday rotation of $\approx$1.5\,rad/m$^2$, comparable to Faraday depth measurements at m-wavelengths, dominated by low-density weakly-magnetized environments.
\end{itemize}
For sake of realism, in these tests, we use the image derived from \cite{Hutschenreuter2024} as dispersion measure and assume a slope of $\gamma=-3$ for the power spectrum of the Galactic magnetic field, in agreement with the findings by the same authors.

Figures containing the results of the inference are shown in Appendix\,\ref{appendixB}. In Figure\,\ref{appendixB1} we show the results for the NVSS-like catalog and in Figure\,\ref{appendixB2} for the LoTSS-like catalog. The simulated Faraday rotation image of our Galaxy used to generate the NVSS-like catalog is the same as shown in the lower right panel of Figure\,\ref{fig1}. Comparing the posterior mean of the Galactic Faraday sky in Figure\,\ref{appendixB1} to the ground truth Galactic Faraday sky in Figure\,\ref{fig1}, we note a high level of fidelity, with larger uncertainties (up to several tens of rad/m$^2$) at the location of the Galactic disk. The larger uncertainty compared to the reconstruction presented in Figure\,\ref{fig4} is likely due to the larger uncertainty of the Faraday depth values in this catalog. 
As for the inference on the extragalactic Faraday rotation, for the NVSS-like catalog, we can recover the values of the parameters $\chi_{\rm red}$ and $\chi_{\rm env}$ within about 1-$\sigma$, while the other two parameters are recovered within about 2-$\sigma$. For the LoTSS-like catalog, i.e. synthetic catalog based on NVSS data for Galactic data points only and on LoTSS data for Faraday rotation values including the extragalactic term, only the parameters describing the redshift and the luminosity dependence $\chi_{\rm lum}$ and $\chi_{\rm red}$ are recovered within the 1-$\sigma$ confidence interval, while the inferred values for the remaining parameters are above 4-$\sigma$.

To investigate above which absolute Galactic latitude the extragalactic recovery is more effective and how this depends on the assumed Galactic magnetic field power spectrum and source density and/or observing wavelength, we run tests considering different cuts in the absolute value of the Galactic latitude of the radio sources used to infer the extragalactic contribution, i.e. $|b|>$0, 20, 45\,degrees, for NVSS- and LoTSS-like catalogs and different slopes of the magnetic field power spectrum. Specifically, we focus on a slope of $\gamma=-3$ and $\gamma=-4$. Indeed, while a slope value of $\gamma=-3$ has been found by \cite{Hutschenreuter2024} to be consistent with the present-day data, a good resemblance with the Galactic Faraday map obtained by \cite{Hutschenreuter2020} can be also obtained for $\gamma=-4$, see \cite{Vacca2026}.
The results are presented in Figure\,\ref{fig6}. The plots indicate that a better recovery of the ground truth values of the parameters is obtained when considering sources away from the Galactic disk. Results for the scenarios $|b|>20\,$degrees and $|b|>45\,$degrees are consistent within the uncertainties, as shown in the first and third column of Figure\,\ref{fig6}. The panels in the second and fourth columns of the same figure show that different cuts in Galactic latitude do not have a significant impact on NVSS-like catalogs. On the contrary, when considering LoTSS-like catalogs, the distance between the inferred and ground truth values is smaller when sources at $|b|>45$\,degrees are considered.
Indeed, in this case, we can constrain all parameters within at most 5-$\sigma$. For NVSS-like catalogs, the results are consistent within the uncertainties when a slope of the Galactic magnetic field power spectrum $\gamma=-3$ (blue dots) and $\gamma=-4$ (red stars) are considered. For LoTSS-like catalogs, the redshift and luminosity dependence are not affected by the different values of the magnetic field power spectrum adopted here, while the intrinsic and environmental terms are better recovered when $\gamma=-4$ is assumed.

We finally note that using broad-band spectro-polarimetric observations, \cite{Ma2019} find that at least 50 out of the 37,543 NVSS polarized radio sources exhibit unreliable Faraday depth values, with a discrepancy on the order of 600\,rad/m$^2$, likely due to the \textit{n}$\pi$-ambiguity, and the corresponding uncertainty in Faraday rotation does not account for this. To investigate the capabilities of our algorithm in extreme cases, as undetected n$\pi$-ambiguities, RM jitter \citep{Stil2025} and, more in general, Faraday complexity phenomena, we run tests with synthetic catalogs including significant deviations in Faraday depths, by adding a quantity reflecting the statistics found by \cite{Ma2019} for different fractions of sources. These tests are presented in Appendix\,\ref{appendixC}. 

\subsection{No extragalactic contribution}
\label{no_eg_contr}
In this section, we explore the possibility that our algorithm can recover an extragalactic contribution even if it is not included in the synthetic catalog generation. To this end, during inference, we considered a one-parameter extragalactic model,
\begin{equation}
\sigma_{\rm eg, i}^2(\Theta)=e^{\chi_{\rm eg}}.
\label{egmodel}
\end{equation}
The rest of the setup is the same described in Sect.\,\ref{functioning}. 
We derive a mean standard deviation of the extragalactic Faraday rotation of approximately 0.27\,rad/m$^2$.  
Note that this value is approximately five times the median of the uncertainty distribution of the Faraday depth adopted here, which is the same as the LoTSS catalog by \cite{OSullivan2023}, see Figure\,\ref{fig3}. 
We investigated if this number is driven by the choice of the prior and our findings suggest that this hypothesis can be excluded, see Appendix\,\ref{appendixD}.

\section{Discussion}
\label{discussion}

We present a new Bayesian algorithm that performs a simultaneous disentangling of the Faraday rotation associated with our own Galaxy from the extragalactic term, properly taking into account the observing noise. The algorithm relies on a catalog of Faraday rotation measurements accompanied by complementary data, as spectroscopic redshifts and radio luminosities, at least for a fraction of the sources.

The algorithm has been tested with synthetic catalogs under different conditions: Galactic magnetic field power spectrum ranging from large to small spatial scales, different cuts in the absolute Galactic latitude of the sources used to infer the extragalactic contribution, uncertainty in Faraday rotation measurements ranging from a few hundredths to about ten rad/m$^2$, variations in the extragalactic Faraday rotation. 

Our findings show that a Galactic magnetic field power spectrum favoring large scales allows for an easier disentangling of the Galactic and extragalactic contribution since, in this case, while the Galactic term is correlated almost only on such scales, the extragalactic term is assumed to be uncorrelated and therefore shows up on small-spatial scales. Good performance is kept up to magnetic field power spectrum slopes of $\gamma=-3$. In particular, when Faraday rotation uncertainties of about 0.06\,rad/m$^2$ are assumed, slopes of $\gamma=-3$ and $\gamma=-4$ ensure a good separation  with a recovery of the extragalactic parameters within at most 2.5-$\sigma$. 

When considering observed-like catalogs, inference suffers of the larger Faraday rotation uncertainty and limited density of polarized sources and/or spectroscopic redshift availability. Furthermore, low- and high-density extragalactic environments  generate different Faraday rotation levels, consequently enabling a recovery of the parameters with different significance.
NVSS-like catalogs have been produced using a standard deviation in the extragalactic Faraday rotation of approximately 10\,rad/m$^2$, mimicing high-density extragalactic media. With this setup, NVSS-like catalogs enable a recovery of all the parameters with a significance between one and two. On the contrary, LoTSS-like catalogs that have been obtained adopting about 1.5\,rad/m$^2$ for the extragalactic Faraday rotation dispersion, better reflect properties of low-density and weakly magnetized environments. In this case, the limited density of polarized sources allows us a recovery with a significance at most of three only for three parameters out of four. 
All these results hold considering a slope of the magnetic field power spectrum $\gamma=-3$, in agreement with \cite{Hutschenreuter2024}, and a distribution of the sources over the sky reflecting the NVSS and LoTSS surveys, meaning no cut in Galactic latitude for the synthetic NVSS-like catalog and only sources with $|b|>$20\,degrees for the synthetic LoTSS-like catalog.

Further tests conducted with different cuts in absolute Galactic latitude and slopes of the Galactic magnetic field power spectrum $\gamma=-3$ and $\gamma=-4$, show better performance when sources with $|b|>20$\,degrees are considered, i.e. away from the Galactic plane. The best results are obtained when only sources above an absolute Galactic latitude of 45\,degrees are included, with inference of the extragalactic parameters at most within 5-$\sigma$. No significant difference is found for NVSS-like catalogs when power spectrum slopes of the magnetic field $\gamma=-3$ and $\gamma=-4$ are used, while slightly better results are found with $\gamma=-4$ for LoTSS-like catalogs. This is not surprising since synthetic LoTSS-like catalogs have been produced assuming a lower extragalactic Faraday rotation standard deviation and a lower density of polarized sources, making it easier to distinguish between the Galactic and extragalactic contributions when the Galactic Faraday rotation shows correlation on larger spatial scales. Overall, these results indicate the applicability of the algorithm to currently available NVSS- and LoTSS-like data in order to constrain the extragalactic Faraday rotation and its dependence on physical quantities such as the redshift of the emitting radio sources. Other statistical approaches have been applied in the past to constrain and characterize extragalactic magnetic fields, however, none of them relies on the simultaneous separation of Galactic and multi-component extragalactic contributions or adequately accounts for observational noise. The approach closest to ours is the one presented by \cite{Oppermann2015}. While the statistical methodology used here has the same theoretical underpinnings as that adopted by \cite{Oppermann2015}, our algorithm is more advanced, as already described in detail in \cite{Hutschenreuter2022}. Furthermore, we note that the discrimination between Galactic and extragalactic contributions proposed in that work is analogous to a one-parameter extragalactic Faraday rotation variance model, whereas in this work we address a multi-component disentanglement. 

Our software appears robust even when spurious contribution to Faraday rotation is included. In order to asses its limits, we also considered the case of contamination of a large number of sources. We still obtain a good inference of the Galactic and extragalactic terms when extragalactic Faraday rotation of about 10\,rad/m$^2$ are considered, meaning that the additional contribution is absorbed by the reconstructed uncertainty through the $\eta$-factors. Inference becomes more difficult for smaller extragalactic contributions. However, good recovery of the redshift and luminosity dependence of the extragalactic contributions is still possible. We expect a similar behavior if the sources used to infer the Galactic Faraday rotation also include an extragalactic contribution, as already demonstrated by \cite{Hutschenreuter2022}.

We can compare these results with the expectations for SKA-Mid and SKA-Low AA4 telescopes Faraday depth catalogs, as presented in \cite{Vacca2026}. At mid-frequencies, using 50,000 extragalactic sources (out of 220,000 sources distributed across the sky, corresponding to approximately a small percentage of those that will be available), we can characterize the contribution to extragalactic Faraday rotation within 1-$\sigma$. In contrast, the NVSS data allow us to constrain the same parameters only within 2-$\sigma$. By considering a low-frequency Faraday depth catalog and using 17,000 sources (out of 210,000
sources distributed across the sky, corresponding to about one third of the Faraday depth measurements that will be available), we can constrain all the extragalactic parameters of our model within about 4-$\sigma$, while with the LoTSS data only three out of four parameters can be constrained in a good way. As shown in the same paper, the SKA telescope data promise a significant step forward in cosmic magnetism; however, they must be complemented by precise location of the sources along the line of sight (i.e., spectroscopic redshifts) to take advantage of the high accuracy of the new generation of radio telescopes, especially at low-frequencies.

We finally investigated how our algorithm performs when no-extragalactic component is included in the synthetic catalog and how different priors affect the results. When no-extragalactic contribution is included in the data but the inference is conducted assuming one extragalactic component, we derive a mean standard deviation of the extragalactic Faraday rotation approximately of 0.27\,rad/m$^2$. This can be interpreted as the minimum value of the extragalactic Faraday rotation that can be derived with this approach. We note that this number is comparable with the median uncertainty in the Galactic reconstruction and gives us an idea of the uncertainties involved in the extragalactic inference.
Since the synthetic catalog used in this test is an ideal catalog including only small Faraday rotation uncertainties, this behavior can be considered as an algorithmic limitation.

\section{Conclusions}
\label{conclusions}
To shed light on the history of cosmic magnetism, the scientific community must detect and characterize large-scale extragalactic magnetic fields. RM-grids are a key tool for this purpose, providing information on all magneto-ionic media encountered by the polarized signal along the line of sight, i.e., Galactic and extragalactic environments. In this context, accurate knowledge of the redshift of the emitting radio sources is crucial, as it allows for precise localization of the source along the line of sight. 

We have developed a new Bayesian algorithm that, given a measurement of the Faraday effect in the direction of a set of point-like radio sources and auxiliary information such as the redshift and the Stokes I luminosity of these sources, is capable to statistically discriminate multiple extragalactic components from the Faraday rotation of our Galaxy, adequately accounting for observational uncertainty. We showed the performance of the algorithm and predictions with synthetic Faraday depth catalogs corresponding to NVSS and LoTSS radio surveys, and demonstrated that the most robust results on the inference of extragalactic parameters are obtained with sources with absolute Galactic latitude greater than 45\,degrees.

Further developments of this algorithm are planned. Knowledge of the large-scale structure of the Universe will allow us to further separate the contribution to Faraday rotation from different extragalactic environments, such as galaxy clusters, filaments, and voids. The density of polarized sources of the new polarization surveys will allow us to study of extragalactic magnetic fields with high spatial resolution. Consequently, it will be necessary to account for the correlated component of these fields. The application of the algorithm to Faraday depth catalogs from current observations is ongoing and will be presented in future work.

\section{Data availability}
Code and numerical results are available on reasonable request.

\begin{acknowledgements}
We thank the referee for the valuable comments and suggestions which helped to improve the manuscript. V. V. acknowledges support from the Prize for Young Researchers "Gianni Tofani" second edition, promoted by INAF-Osservatorio Astrofisico di Arcetri (DD n. 84/2023).
S.\ H.\ acknowledges funding by the European Union (ERC, ISM-FLOW, 101055318). Views and opinions expressed are, however, those of the author(s) only and do not necessarily reflect those of the European Union or the European Research Council. Neither the European Union nor the granting authority can be held responsible for them. 
J. R. acknowledges financial support from the German Federal Ministry of
Education and Research (BMBF) under grant 05A23WO1 (Verbundprojekt
D-MeerKAT III). This work was carried out thanks to the funding of the Regione Autonoma della Sardegna, ai sensi della Legge Regionale 7 agosto 2007, n.7 "Promozione della Ricerca Scientifica e dell'Innovazione Tecnologica in Sardegna".
\end{acknowledgements}

\bibliographystyle{aa} 
\bibliography{bibliography.bib}

\begin{appendix}

\section{Model}
\label{appendixA}

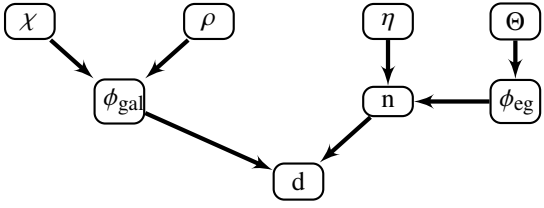
\begin{figure}[h!]
\begin{tikzpicture}[auto]
  \matrix[ampersand replacement=\&, row sep=0.5cm, column sep=0.5cm] {
   \&
    \node [block] (chi) {$\chi$};
    \&
    \&
    \node [block] (rho) {$\rho$};
    \&
    \&
   \node [block] (eta) {$\eta$};
    \&
    \&
 \node [block] (theta) {$\Theta$};
    \&
    \\
   \&
    \&
    \node [block] (phi) {$\phi_{\rm gal}$};
    \&
    \&
    \&
   \node [block] (n) {$n$};
    \&
    \&
   \node [block] (phi_eg) {$\phi_{\rm eg}$};
    \&
    \\
    \&
    \&
    \&
    \&
    \node [block] (d_phi) {$d$}; 
    \&
    \&
    \&
    \&
    \\ 
    }; 
    \path [line] (chi) -- node{}(phi);
    \path [line] (rho) -- node{}(phi);
    \path [line] (phi) -- node{}(d_phi);
    \path [line] (n) -- node{}(d_phi);
    \path [line] (eta) -- node{}(n);
    \path [line] (theta) -- node{}(phi_eg);
    \path [line] (phi_eg) -- node{}(n);

\end{tikzpicture}
\caption{Graphical representation of the hierarchical Bayesian model adopted in this work. The observed data \textit{d} are the result of the sum of the Galactic Faraday depth and noise. The Galactic Faraday depth $\phi_{\rm gal}$ depends on the fields $\chi$ and $\rho$ via Eq.\,\ref{galcontr}. The noise \textit{n} is assumed to be drawn from a zero-mean Gaussian. For Galactic points only, see Eq.\,\ref{galonly}, the noise variance is corrected by multiplying it by unknown correction factors $\eta$. For extragalactic data points, see Eq.\,\ref{galplusegal}, the observed noise variance is added in quadrature to the variance of the extragalactic Faraday rotation, which is a function of the parameters $\Theta$, $\sigma_{\rm eg}^2=\sigma_{\rm eg}^2(\Theta)$. 
}
\label{appendixA1}
\end{figure}

Graphical representation of the hierarchical Bayesian model adopted in this work is shown in Figure\,\ref{appendixA1}.  

\section{Inference for observed-like catalogs}
\label{appendixB}

\begin{figure}[h!]
\centering
\includegraphics[width=\linewidth]{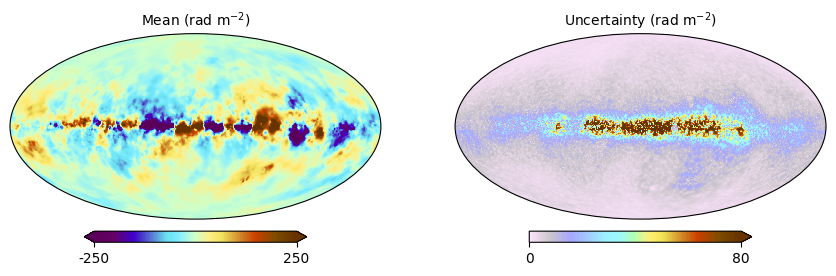}\\
\vspace{0.5cm}
\includegraphics[width=\linewidth]{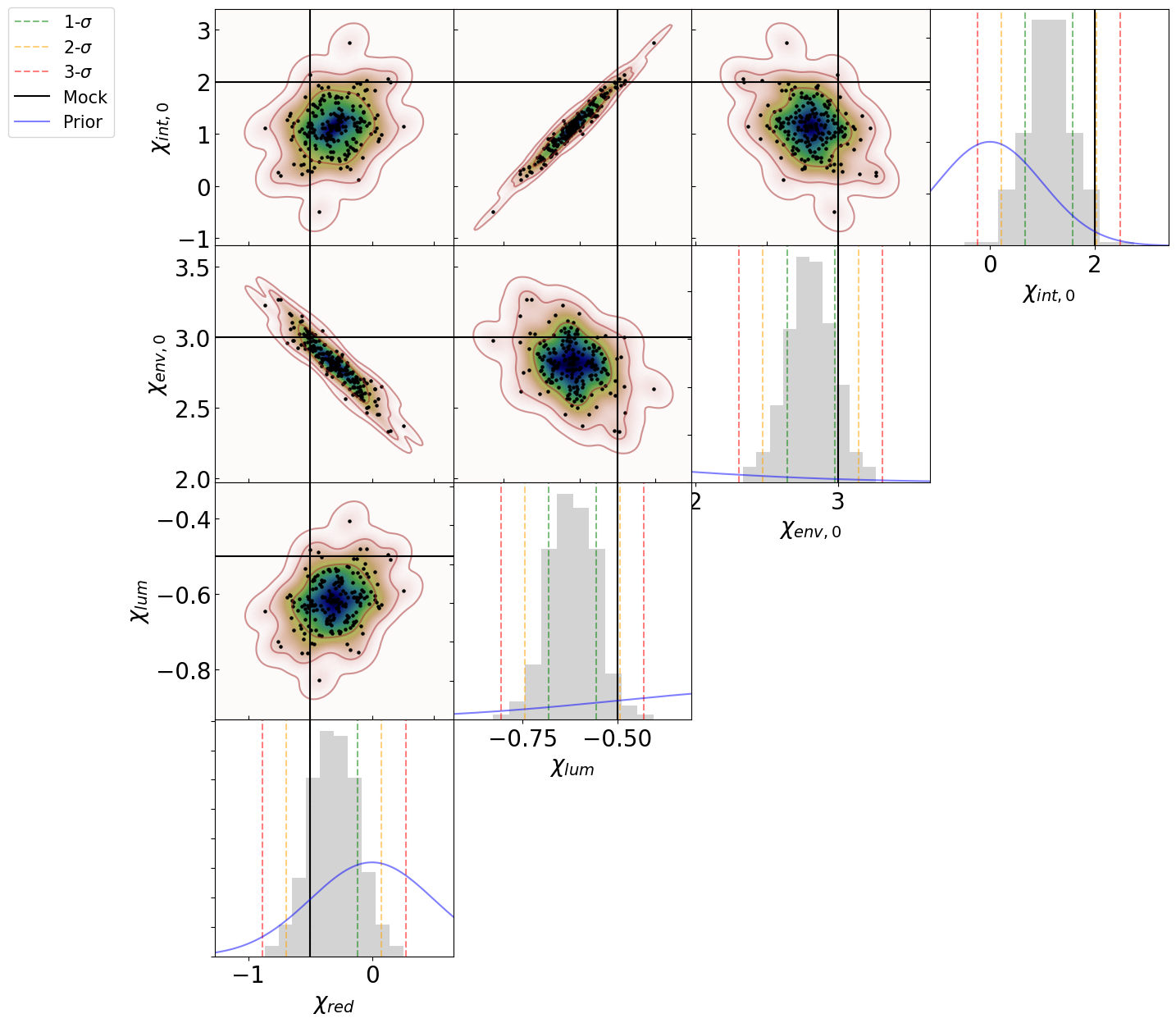}
\caption{As in Figure\,\ref{fig4} but for a NVSS-like catalog.}
\label{appendixB1}
\end{figure}

\begin{figure}[h!]
\centering
\includegraphics[width=\linewidth]{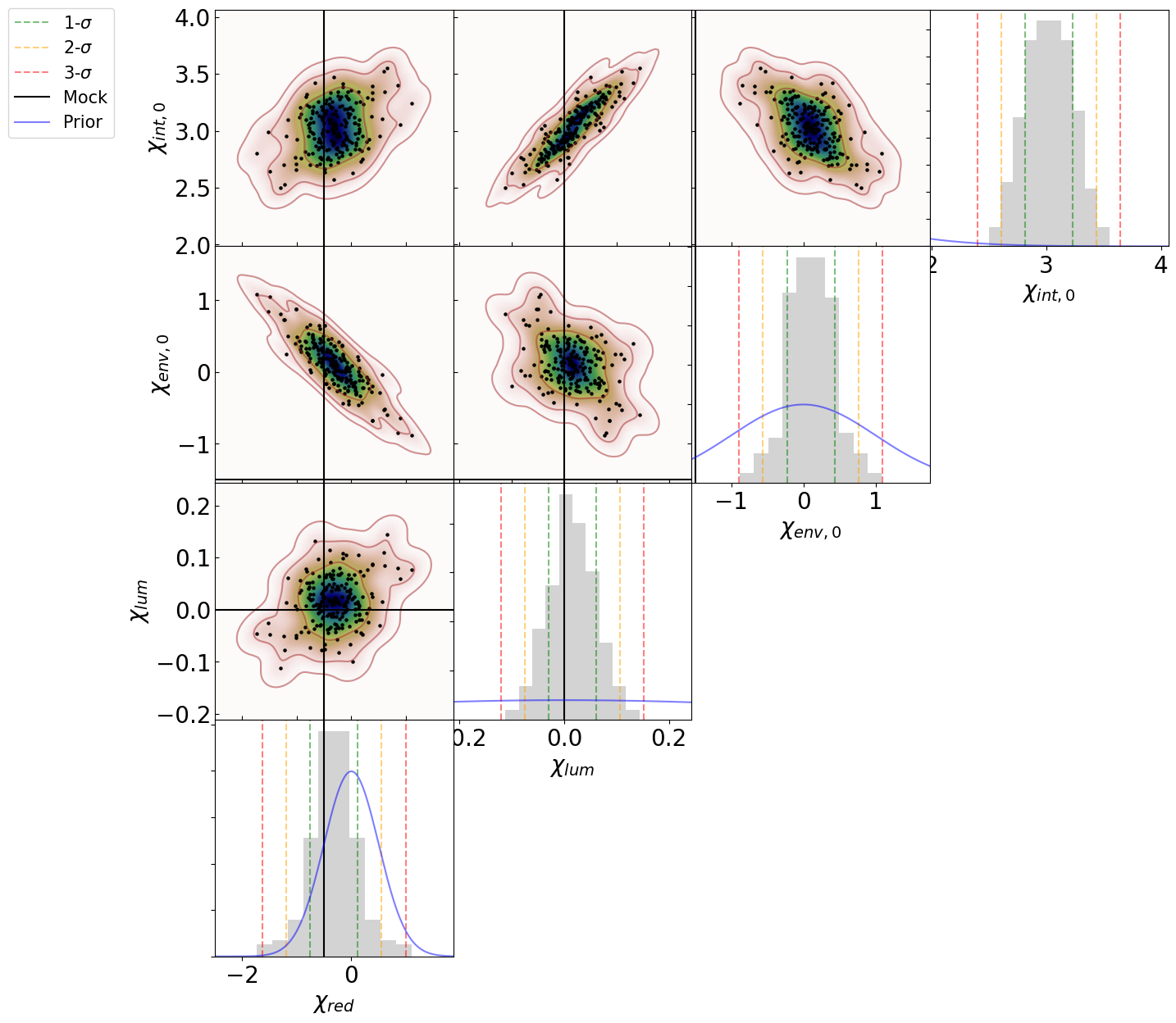}
\caption{As in the bottom panel of Figure\,\ref{fig4} but for a LoTSS-like catalog.}
\label{appendixB2}
\end{figure}

In Fig.\,\ref{appendixB1} and  in Fig.\,\ref{appendixB2} we present the result of the inference for NVSS- and LoTSS-like catalogs respectively, as described in Sect.\,\ref{etafactor}.

\section{Stress-tests} 
\label{appendixC}
In this Appendix we present stress-tests using synthetic catalogs produced following the prescriptions for NVSS- and LoTSS-like in Sect.\ref{etafactor}. 
In order to check the robustness of the algorithm with respect to extreme cases as those related to possible undetected n$\pi$-ambiguities \citep[e.g.,][]{Ma2019}, rotation measure jitter \citep{Stil2025} and, more in general, Faraday complexity, we randomly add/subtract for a fraction of the data points a spurious contribution. This contribution has been drawn from a Gaussian with mean 605.0\,rad/m$^2$ and standard deviation 65.0\,rad/m$^2$, following the findings by \cite{Ma2019}. 
In our setup, without further assumptions, this additional term can mimic both sources with low real Faraday rotation but high measured Faraday rotation, and sources with high real Faraday rotation but low measured Faraday rotation. 
To better investigate the limits of our algorithm, we have produced synthetic catalogs corresponding to a different number of sources \textit{M} that exhibit this problem: 
\begin{itemize}
\item \textit{M}=50, the minimum number identified by \cite{Ma2019};
\item \textit{M}= 5000, about 10-15\,percent of NVSS sources; 
\item \textit{M}= 25000, most NVSS sources ($\gtrsim$50\,percent).
\end{itemize}
This spurious term has been included only for sources among those used to infer the Galactic Faraday sky only. For this class of  sources, a multiplicative $\eta$ factor has been implemented in the noise variance, which is inferred from the data, see description in Sect.\,\ref{methodology}.

\begin{figure}[h!]
\centering
\includegraphics[width=\linewidth]{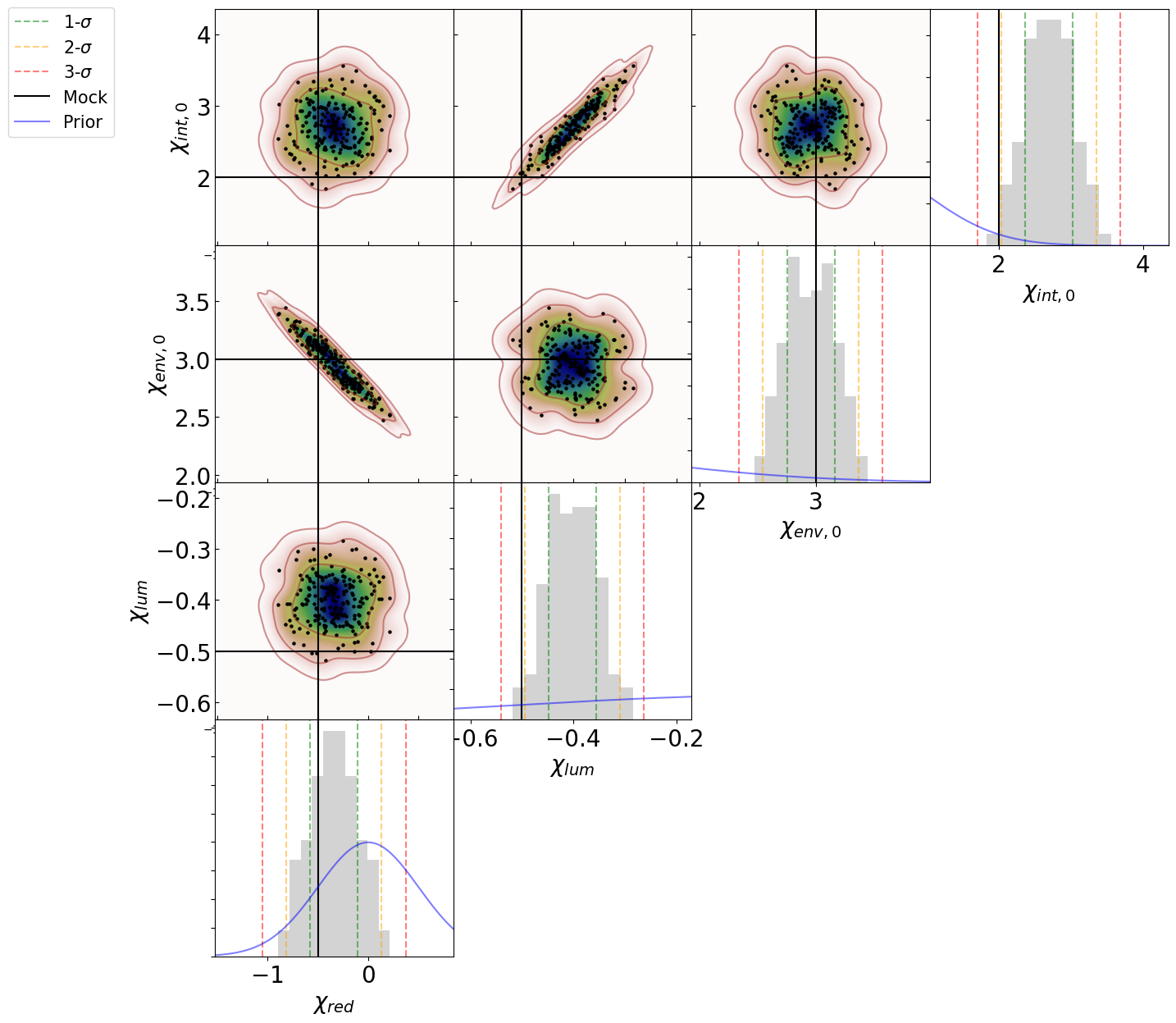}
\caption{As in bottom panel of Figure\,\ref{fig4} but for a NVSS-like catalog with n$\pi$-ambiguities for 5000 sources.}
\label{appendixC1}
\end{figure}

\begin{figure}[h!]
\centering
\includegraphics[width=\linewidth]{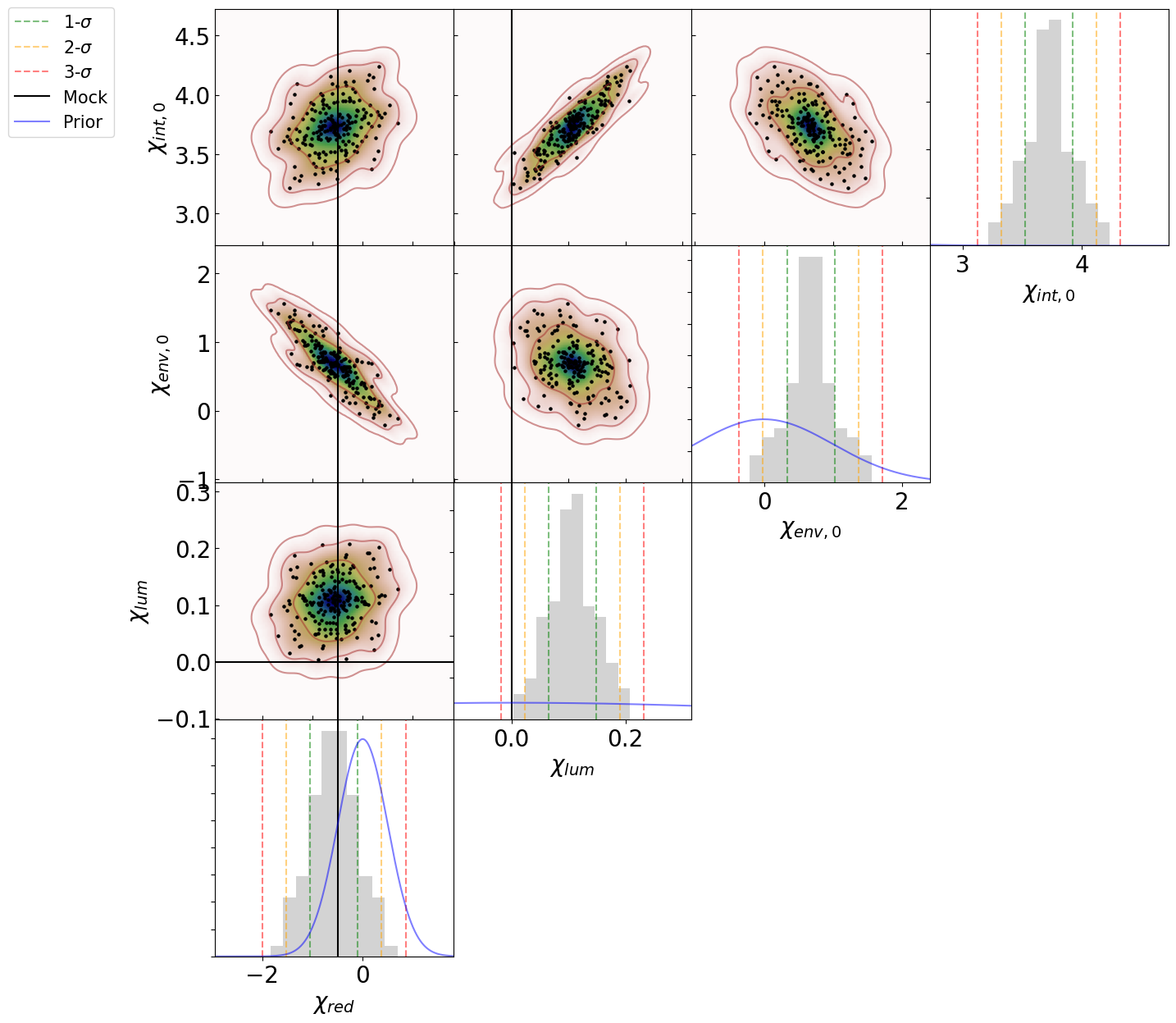}
\caption{As in bottom panel of Figure\,\ref{fig4} but for a LoTSS-like catalog with n$\pi$-ambiguities for 5000 sources.}
\label{appendixC2}
\end{figure}
\begin{figure}
\includegraphics[width=\linewidth]{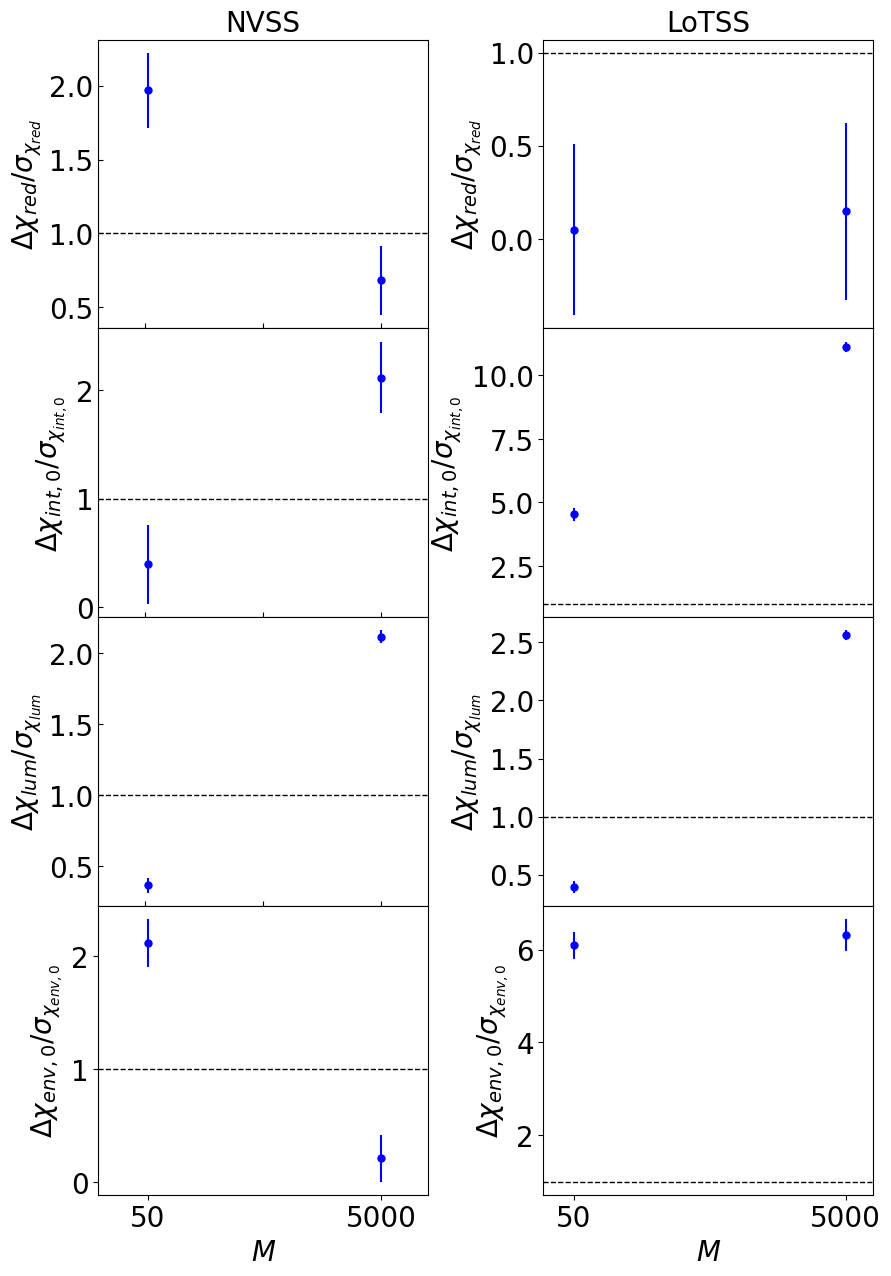}
\caption{Difference between the inferred value and the ground truth value in units of its a posteriori uncertainty of 1-$\sigma$  as a function of $M$ for $\chi_{\rm red}$, $\chi_{\rm int,0}$, $\chi_{\rm lum}$, and $\chi_{\rm env, 0}$, from top to bottom respectively. The horizontal dashed line in each panel shows a reference value of 1. Left panels show results for NVSS-like catalogs, while right panels for LoTSS-like catalogs.}
\label{appendixC3}
\end{figure}

In Figure\,\ref{appendixC1} we show the results for a NVSS-like and in Figure\,\ref{appendixC2} for a LoTSS-like catalog, considering \textit{M}=5000. In Figure\,\ref{appendixC3}, we compare the results for both NVSS- and LoTSS-like catalogs for \textit{M}=50 and \textit{M}=5000. 
For the NVSS-like catalog, i.e. synthetic catalog based only on NVSS data, the results are comparable with the findings in Sect.\,\ref{etafactor}. Furthermore, for \textit{M}=50 and \textit{M}=5000, the results are comparable and all parameters are recovered within 2-$\sigma$, i.e. the number of data points affected by the \textit{n}$\pi$-ambiguity is so small that it does not have a significant impact on the inference of the extragalactic Faraday rotation.  On the contrary, for the LoTSS-like catalog, the algorithm has more difficulties in recovering the parameters and the results are slightly worst when considering \textit{M}=5000.
For the most extreme case, \textit{M}=25000, both for NVSS-like and for LoTSS-like catalogs, the results are unreliable. Indeed, we observe strong artifacts in the Galactic Faraday sky, mainly on small spatial scales. This is not surprising, since in this case about 50\,percent of the data is affected by an undetected \textit{n}$\pi$-ambiguity. 
Howover, the Galactic Faraday image obtained by \cite{Hutschenreuter2022} and, previously, by \cite{Oppermann2015} shows no obvious artifacts, but a smooth reconstruction, suggesting that such a large fraction of data points is not affected by this issue.

The results presented in this Appendix indicate that our algorithm is robust and shows good performance also in extreme cases where even $\sim$10-15\,percent of the Faraday rotation measurements significantly differ from the real values.

\section{Influence of the prior} 
\label{appendixD}
\begin{figure}
\includegraphics[width=\linewidth]{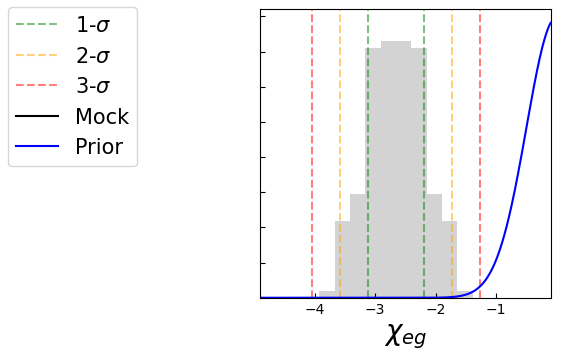}
\includegraphics[width=\linewidth]{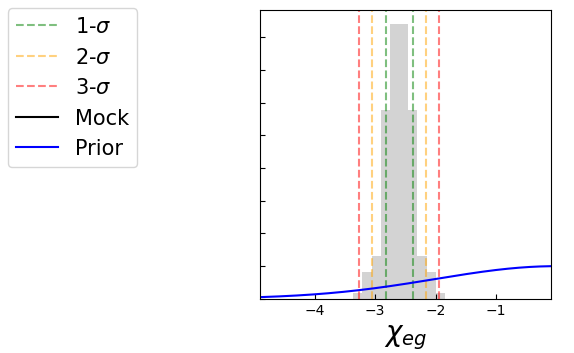}
\includegraphics[width=\linewidth]{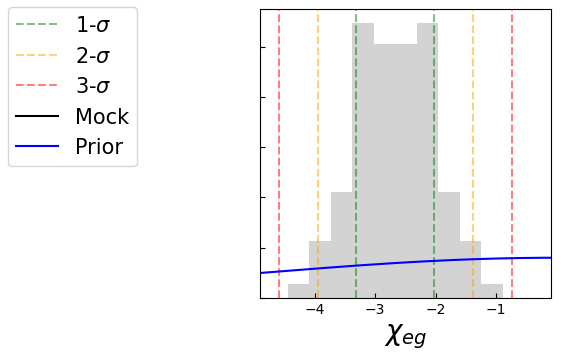}
\caption{Posterior distribution obtained by assuming no extragalactic Faraday rotation in the generation of the synthetic catalog, but a one-parameter extragalactic model during inference. The green, orange, and red dashed lines indicate the 1-, 2-, and 3-$\sigma$ of the posterior distribution. The blue line shows the prior distribution. 
The top, middle, and bottom panel refer respectively to a a zero-mean Gaussian prior with standard deviation equal to  0.5, 2, and 5.}
\label{appendixD1}
\end{figure}

In this Appendix we try to address the impact of the prior choice on the results presented in Sect.\,\ref{no_eg_contr}, 
i.e. whether smaller extragalactic contributions are excluded because penalized by the prior. To this end, we adopt three different priors, i.e. a zero-mean Gaussian with standard deviation equal respectively to 0.5, 2.0 and 5. In these cases, we derive a mean standard deviation of the extragalactic Faraday rotation respectively of 0.27$\pm$0.06\,rad/m$^2$, 0.27$\pm$0.03\,rad/m$^2$, and 0.26$\pm$0.08\,rad/m$^2$.
Our results, shown in Figure\,\ref{appendixD1}, indicate that the inferred parameter is consistent within the uncertainty when these three values are considered for the prior standard deviation, suggesting that the prior choice is not affecting our findings and the inferred values are rather driven from the information encoded in the data.

\end{appendix}
\end{document}